\documentclass[letter,11pt]{article}
\usepackage{graphicx}

\def\be{\begin{equation}}
\def\ee{\end{equation}}
\def\bea{\begin{eqnarray}}

\def\eea{\end{eqnarray}}

\begin{document}

\title{BEC Collapse and Dynamical Squeezing of Vacuum Fluctuations}
\author{E. A. Calzetta$^1$\thanks{%
Email: calzetta@df.uba.ar} and B. L. Hu$^2$\thanks{%
Email: hub@physics.umd.edu} \\
$^1${\scriptsize Departamento de Fisica, Facultad de Ciencias Exactas y
Naturales,}\\
{\scriptsize Universidad de Buenos Aires- Ciudad Universitaria, 1428 Buenos
Aires, Argentina}\\
$^2${\scriptsize Department of Physics, University of Maryland, College
Park, MD 20742, USA}}
\date{{\small (May 18, 2003, second version of cond-mat/0207289 deposited in arXiv
on July 11, 2002. umdpp02-61)}}
\maketitle

\begin{abstract}
We analyze the phenomena of condensate collapse, as described by Donley et
al \cite{JILA01b,Claussen03}, by focusing on the behavior of excitations or
fluctuations above the condensate, as driven by the dynamics of the
condensate, rather than the dynamics of the condensate alone or the kinetics
of the atoms. The dynamics of the condensate squeezes and amplifies the
quantum excitations, mixing the positive and negative frequency components
of their wave functions thereby creating particles which appear as bursts
and jets. By analyzing the changing amplitude and particle content of these
excitations, our simple physical picture explains well the overall features
of the collapse phenomena and provide excellent quantitative fits with
experimental data on several aspects, such as the scaling behavior of the
collapse time and the amount of particles in the jet. The predictions of the
bursts at this level of approximation is less than satisfactory but may be
improved on by including the backreaction of the excitations on the
condensate. The mechanism behind the dominant effect -- parametric
amplification of vacuum fluctuations and freezing of modes outside of
horizon -- is similar to that of cosmological particle creation and
structure formation in a rapid quench (which is fundamentally different from
Hawking radiation in black holes). This shows that BEC dynamics is a
promising venue for doing `laboratory cosmology'.
\end{abstract}

\newpage

\section{Introduction}

We introduce a new perspective in the analysis of the phenomena of
condensate collapse, described by Donley et al \cite{JILA01b,Claussen03}, by
focusing on the behavior of fluctuations above the condensate, rather than
the condensate itself. We show that the condensate dynamics squeezes,
amplifies, and mixes positive and negative frequency components of the wave
functions of the condensate excitations. In addition to providing a good
qualitative understanding of the general picture our theory also produces
precise predictions, specifically, on the critical number of particles at
the first instance when the instability sets in, the scaling of the waiting
time $t_{collapse}$ and the number of particles in a jet. In this rendition
we point out the analogy between the evolution of quantum excitations of the
collapsing condensate and the vacuum fluctuations parametrically amplified
by the background spacetime in the Early Universe, suggesting a new venue
for ``laboratory cosmology''.

A condensate formed from a gas of cold ($3$nK) Rubidium atoms is rendered
unstable by a sudden inversion of the sign of the interaction between atoms.
After a waiting time $t_{collapse},$ the condensate implodes, and a fraction
of the condensate atoms are seen to oscillate within the magnetic trap which
contains the gas (see below and \cite{JILA01b}). These atoms are said to
belong to a ``burst''. In the experiments described by Donley et al. -- to
single out this controlled BEC collapse experiment from the others, we shall
adhere to the namesake Bose Novae -- the interaction is again suddenly
turned off after a time $\tau_{evolve}.$ For a certain range of values of $%
\tau _{evolve},$ new emissions of atoms from the condensate are observed,
the so-called ``jets''. Jets are distinct from bursts: they are colder,
weaker, and have a distinctive disk-like shape.

The Donley et al. experiment takes full advantage of the tunability of the
effective atomic interaction due to a Feshbach resonance characteristic of $%
^{85}$Rubidium \cite{JILA98,JILA00}. The resonance is caused by the presence
of a bound state whose binding energy may be tuned by means of an external
magnetic field. In later experiments \cite{JILA02a,JILA02b}, observed
fluctuations in the number of particles in the condensate have been
well-explained as arising from oscillations between the usual atomic
condensate and a molecular state \cite
{KGB02,KH02,MSJ02a,MSJ02b,Mackie02,Yin03}.

These oscillations were observed for magnetic fields in the order of $160$G,
where the effective scattering length is of the order of $500a_{0}$ (and
positive) ($a_{0}=0.529$\ $10^{-10}$m\ \ \ is the Bohr radius) and the
frequency of oscillations is of hundreds of KHz \cite{JILA02a,JILA02b}. By
contrast, in the Donley et al. experiment \cite{JILA01b}\ typical fields
were around $167$G, the scattering length was only tens of Bohr radii (and
negative) and the frequency of atom - molecule oscillations may be estimated
as well over ten MHz \cite{JILA03}. Under these conditions it is unlikely
that the molecular condensate plays any important dynamical role, and indeed
no oscillations are reported in the original paper (for the opposite view,
see \cite{MMH03}). For these reasons and to highlight the mechanism
particular to this experiment, we shall not include explicitly a molecular
condensate in our model but discuss in detail the one - field model.
However, if need to, this may be done in a very simple way, by including a
second field to describe molecular destruction and creation operators \cite
{MSJ02a,MSJ02b,MMH03}. We will elaborate on this point in a later subsection

There is a vast literature attempting to provide theoretical explanations of
collapsing condensates \cite{SHU96,SSH98,KMS98,BR00,HUA01,AD02,SU03}. In
addition to speculations that Bose Novae is due to molecular oscillations as
alluded above (which we view as of secondary importance) the most serious
theoretical attempt is based on the Gross-Pitaevskii equation with
explicitly introduced nonlinear terms to account for multiparticle
interactions \cite{SU03,SRH02}. We will show that the primary mechanism
responsible for the main features of the Bose Novae experiment originates
from the dynamics of quantum fluctuations around the background condensate
field(s). We start with the Heisenberg operator for the many body wave
function and split it into a c-number part describing the condensate
amplitude and a q-number part describing collective excitations (not
individual atoms) above the condensate. We then derive an evolution equation
for the wave function operator of the quantum (non-condensate) excitations
under an improved Hartree approach, the so-called Popov approximation \cite
{POP87,GRI93,GRI96,PS02}.

In this paper, we use a ``test field'' approximation, by adopting (rather
than deriving) the specific evolution of the condensate extracted from the
experiment as given and study the dynamics of the excitations riding on this
dynamics. Note that the experimentally given condensate dynamics is
different from the mean field dynamics obtained from a solution of the GP
equation, because the former includes the dynamical effects of the
fluctuations. Finding a self-consistent solution of the evolution equations
for both the condensate and its fluctuations is called the `backreaction
problem'. It has been studied in detail in problems of similar contexts such
as cosmological particle production (see below). Theoretical investigations
for BEC fluctuations dynamics can be found in Refs. \cite{KGB02,KB02,DS02}.
The squeezing of quantum unstable modes and its back reactions on the
condensate has been considered before, e.g., as a damping mechanism for
coherent condensate oscillations \cite{KM01}, and also applied to the
description of condensate collapse \cite{VM02,Y02,VYA01,YBJ02,R76}. Field
theory methods have recently been applied to the problem of formation and
stability of Bose condensates \cite{field,BWLYA01}. Fluctuations have also
been considered by G\'{o}ral et. al. \cite{GGR00} and Graham et al. \cite
{GWCFW96}. Our work differs from them in the emphasis we place on the
behavior of the quantum excitations as a consequence of condensate dynamics.

Particularly relevant to the present work is Ref. \cite{Y02}, where condensate 
collapse is analyzed from the point of view that 
the physics is mainly due to the dynamics of 
quantum fluctuations, the same view as we hold here. There, the trapping 
potential is replaced by a normalizing box, whose volume is eventually 
taken to infinity. Our analysis in Sections 2 and 3 is 
for a more realistic geometry, which enables us to compare quantitatively to experiments. The analysis of bursts versus jets given in Section 4 however originates from a new concept inspired by cosmological processes. 

To the extent that many phenomena observed in connection to the collapse of
this nature (Bose Novae) are essentially the result of a quantum fluctuation
field (the non-condensate) interacting with a time dependent background (the
condensate), as we believe it is, there is a close analog with similar
processes in the early universe, specifically, vacuum particle creation from
an time dependent external field \cite{Schwinger} or in a curved background
spacetime \cite{Parker,Zeldovich}. (For a squeezed state depiction of this
process, see, e.g., \cite{HKM} and references therein.) One could view
condensate collapse as a laboratory realization of cosmological particle
creation during quenching. (Note this is not the physical process behind
black hole particle creation, as in the Hawking effect, much attention drawn
to its detection in BEC notwithstanding \cite{HawEffBEC}.) In this process
there is a competition between two (inverse) time scales, the physical
frequency of the mode under consideration, and the inverse collapse
(expansion) rate of the condensate. In cosmology the inverse expansion rate
is the Hubble constant for the background spacetime. While a mode whose
physical frequency is higher than the Hubble constant, we refer to it as
``inside the horizon'', and its behavior is oscillatory. When the converse
obtains, the mode is ``outside the horizon''. They are depicted as `frozen'
because they do not oscillate (see below), but are amplified \cite{STA86}.
This amplification is largely responsible for the observed primordial
density contrast in the Universe \cite{BM}.

In the Bose-Novae collapse problem, the role of the ``Hubble'' constant is
played by the inverse growth (exponential) rate of the most unstable mode of
the condensate, which is determined by the instantaneous number of particles
in the condensate. Modes whose natural frequency is greater than the
corresponding scale are relatively impervious to the dynamical condensate,
but when the converse obtains, consequences are drastic. When the
exponential growth is the dominant factor, the mode is frozen; instead of
oscillating, it is being amplified, a process which is analogous to the
growth of fluctuations during spinodal decomposition \cite{SD}.

In the same way that modes that left the horizon during inflation return
during the radiation and matter dominated eras, giving rise to acoustic
oscillations, as the unstable condensate sheds its atoms and approaches
stability, the band of ``frozen'' modes narrows: we say that modes ``thaw''
as they turn from exponentially increasing to oscillatory behavior. The crux
of the matter is that only oscillating modes are detected through
destructive absorption imaging (see below). Whenever a mode thaws, it is
perceived as if particles were being created. In the conditions of the
experiment the initial number of actual particles above the condensate is
negligible, and we may describe the phenomenon as particle creation from the
vacuum.

To summarize, the key idea in our understanding of the phenomena associated
with a condensate collapse is that of a dynamical background field of the
condensate squeezing and mixing the positive and negative frequency
components of its quantum excitations, thereby creating particles from the
vacuum. The viewpoint of this work may be easily incorporated in a first
principles approach as taken in e.g. \cite{STO99,DS01b,GZ00,GAF01}. The
remarkable analogy between condensate collapse and quantum processes in the
Early Universe and spinodal decomposition in phase transitions may stimulate
new related experiments in BEC to be carried out to address these problems
in cosmology and condensed matter physics \cite{other}.

This paper is organized as follows. Section II we briefly review the
phenomenology of condensate collapse and set up the basic mathematical
model. In Section III we give a discussion of the onset of instability and
of the scaling of the waiting time $t_{collapse}$ with the scattering
length. In Section IV, we turn to a discussion of bursts and jets, based on
the distinction between frozen and thawed modes. By postulating a specific
condensate evolution (extracted from the experiment) we obtain quantitative
predictions for the number of particles in a jet as a function of the time $%
\tau _{evolve}$ (when the scattering length is brought to zero). Our results
are summarized in the final Section. A few technical details are left to the
Appendixes.

\section{The model}

\subsection{An overview of condensate collapse}

In the Bose Novae experiment \cite{JILA01b} a gas of $N$ $^{85}$Rb atoms at
a temperature of $3$ nK is prepared in a state where they behave essentially
as a free gas within an anisotropic harmonic trap (see Figs. 1 and 2). The
trap has a cylindrical geometry [let $\mathbf{r}\equiv \left( \rho ,\varphi
,z\right) $ be the usual cylindrical coordinates, with $\varphi $ the
azimuthal angle]; the trap frequencies are $\nu _{radial}=17.5$ Hz ($\omega
_{radial}=2\pi \nu _{radial}=110$ Hz) and $\nu _{axial}=6.8$ Hz ($\omega
_{axial}=42.7$ Hz). At time $t=0,$ the scattering length $a$ (see below) is
suddenly turned to a negative value $-a_{collapse}$. This configuration is
known to be unstable whenever the number $N_{0}$ of atoms in the condensate
exceeds $\kappa a_{ho}/\left| a\right| $, where $a_{ho}$ is a characteristic
length of the trap. The coefficient $\kappa $ was reported as $\kappa =0.46$ 
\cite{JILA01}, but later measurements suggest it should be raised to $\kappa
=0.55$ \cite{Claussen03,JILA03}.

For $N_{0}=6000$ atoms, the instability threshold is reported at $\left|
a\right| =a_{cr}=5.12\;a_{0}$, where $a_{0}=0.529$\ $10^{-10}$m. Therefore $%
a_{ho}=5.6\;10^{4}a_{0}=3$\ $10^{-6}$m. If we write $a_{ho}^{2}=\hbar
/M\omega $ and introduce the atomic mass of $85m_{P}\sim 1.3\;10^{-25}$kg
and $\hbar =1.05\;10^{-34}$Js$,$ this corresponds to a frequency $\omega =90$%
Hz. This agrees well with the geometric average $\omega =\left( \omega
_{axial}\omega _{radial}^{2}\right) ^{1/3}=80$Hz. It follows that $%
t_{ho}=\omega ^{-1}=12.5$ms.

In spite of the instability, over a time $t_{collapse}$ there is no
significant decay of the condensate. $t_{collapse}$ depends very strongly on 
$a_{collapse}$ (see below). After $t_{collapse},$ the number of atoms in the
condensate falls exponentially$.$ If left to itself, the condensate
eventually stabilizes retaining a number $N_{remnant}$ of atoms.

During this period, a cloud or burst of atoms is observed. These atoms
oscillate in the radial direction with the trap radial frequency. The energy
associated with the burst is larger in the radial direction than in the
axial direction. The number of atoms in the burst increases with the time
elapsed since the decay begins, with a time constant of about $1.2$ ms;
after $7$ ms, the number of burst atoms stabilizes. About a fifth of the
atoms in the condensate go into the burst, with variations of about $20$\%.

Condensate decay is interrupted at a time $t=\tau _{evolve}$, when the
scattering length is again tuned to a positive value. If the condensate is
already stabilized at $\tau _{evolve},$ nothing too drastic happens, but
otherwise a new phenomenon appears, namely the expulsion of a jet of atoms.
Jets have much lower kinetic energy than bursts (a few nano-Kelvin).

\subsection{Basic equations\protect\footnote{%
In writing down the main equations of our model, we choose a sign convention
which makes the effective coupling constant positive for an attractive
interaction, and a system of units adapted to the problem. Taking the
average frequency $\omega $ as reference, we may define length and time
scales $a_{ho}$ and $t_{ho}$ (see above) and an energy scale $E_{ho}=\hbar
\omega =M\omega ^{2}a_{ho}^{2}$. From now on, we shall choose units such
that these three scales take the value $1$.}}

The model is based on the Hamiltonian operator for N interacting atoms with
mass M in a trap

\begin{equation}
\mathbf{\hat{H}} = \int d\mathbf{r}\;\left\{ -\frac{\hbar ^{2}}{2M}\mathbf{%
\Psi }^{\dagger }\nabla ^{2}\mathbf{\Psi }+\mathbf{V}\left( \mathbf{r}%
\right) \mathbf{\Psi }^{\dagger }\mathbf{\Psi }-\frac{U}{2}\mathbf{\Psi }%
^{\dagger 2}\mathbf{\Psi }^{2}\right\}
\end{equation}
with the total number operator $N$ given by

\begin{equation}
N=\int d\mathbf{r}\;\mathbf{\Psi }^{\dagger }\mathbf{\Psi }
\end{equation}
Here $\mathbf{V}\left( \mathbf{r}\right) $ is the trap potential and $U$ is
(assumed to be a short ranged) the interaction between the atoms. We
introduce a dimensionless field operator $\Psi \left( x\right) $

\begin{equation}
\mathbf{\Psi }\left( r\right) \equiv a_{ho}^{-3/2}\Psi \left( x\right)
\end{equation}
where $a_{ho}$ is a characteristic length of the trap, and a dimensionless
coupling constant $u$

\begin{equation}
U \equiv \hbar \omega a_{ho}^{3}u
\end{equation}

In terms of the scattering length $a$ (which we define as positive for an
attractive interaction), we have

\begin{equation}
U=\frac{4\pi \hbar ^{2}a}{M}
\end{equation}
So

\begin{equation}
u=\frac{4\pi \hbar a}{M\omega a_{ho}^{3}}=4\pi \left(\frac{a}{\,\,\,\,a_{ho}}%
\right)
\end{equation}

The Hamiltonian and the trap potential can also be written in terms of
dimensionless variables 
\begin{equation}
\mathbf{\hat{H}=}E_{ho}\hat{H}, \ \ \mathbf{V}\left( \mathbf{r}\right)
=E_{ho}V\left( r\right)
\end{equation}
Assuming a cylindrical shaped potential

\begin{equation}
V\left( \mathbf{r}\right) =\frac{1}{2}( \omega _{z}^{2}z^{2}+\omega _{\rho
}^{2}\rho ^{2})
\end{equation}
with radial $\rho$ and longitudinal $z$ coordinates measured in units of $%
a_{ho,}$ with associated (dimensionless) frequencies $\omega _{z}=\omega
_{axial}/\omega \sim 1/2$ and $\omega _{\rho }=\omega _{radial}/\omega \sim 
\sqrt{2}.$

$\Psi $ obeys the equation of motion

\begin{equation}
\dot{\Psi}=i\left[ \hat{H},\Psi \right]
\end{equation}
and the equal time commutation relations (ETCR)

\begin{equation}
\left[ \Psi \left( t,\mathbf{r}\right) ,\Psi ^{\dagger }\left( t,\mathbf{%
r^{\prime}}\right) \right] =\delta ^{\left( 3\right) }\left( \mathbf{r}-%
\mathbf{r^{\prime}}\right),  \label{EomPsi}
\end{equation}
whereby

\begin{equation}
i\frac{\partial \Psi }{\partial t}=H\Psi -u\Psi ^{\dagger }\Psi ^{2}.
\end{equation}
Here

\begin{equation}
H=-\frac{1}{2}\nabla ^{2}+V\left( r\right)  \label{hachetrap}
\end{equation}
is the (dimensionless) one-particle trap Hamiltonian. We decompose the
Heisenberg operator $\Psi $ into a c-number condensate amplitude $\Phi $ and
a q-number noncondensate amplitude $\psi$, consisting of the fluctuations or
excitations

\begin{equation}
\Psi (\mathbf{r},t)=\Phi (\mathbf{r},t)+\psi (\mathbf{r},t)
\end{equation}
Then 
\begin{equation}
\Psi ^{\dagger }\Psi =|\Phi |^2+\Phi (\psi ^{\dagger }+\psi )+\psi ^{\dagger
}\psi
\end{equation}
and 
\begin{equation}
n\equiv <\Psi ^{\dagger }\Psi >=|\Phi |^2+<\psi ^{\dagger }\psi >\equiv n_0+{%
\tilde n,}
\end{equation}
where $\left\langle {}\right\rangle $ denotes expectation value with respect
to the initial quantum state. The cross terms vanish because the mean field
is defined with zero fluctuations. Here $n_0=\left| \Phi \right| ^2$ is the
number density of the condensate and ${\tilde n}\equiv \left\langle \psi
^{\dagger }\psi \right\rangle $ is the number density corresponding to the
fluctuations. Further,

\begin{eqnarray}
\Psi ^{\dagger }\Psi ^{2} &=&\left( \Phi ^{\ast }+\psi ^{\dagger }\right)
\left( \Phi ^{2}+2\Phi \psi +\psi ^{2}\right)  \nonumber \\
&=&\left| \Phi \right| ^{2}\Phi +2\left| \Phi \right| ^{2}\psi +\Phi
^{2}\psi ^{\dagger }+2\Phi \psi ^{\dagger }\psi +\Phi ^{\ast }\psi ^{2}+\psi
^{\dagger }\psi ^{2}
\end{eqnarray}

Under the self-consistent mean field approximation \cite{GRI96}

\begin{equation}
\psi ^{\dagger }\psi \simeq \left\langle \psi ^{\dagger }\psi \right\rangle =%
{\tilde{n}}
\end{equation}

\begin{equation}
\psi ^{2}\simeq \left\langle \psi ^{2}\right\rangle ={\tilde{m}}
\end{equation}
where ${\tilde{m}}$ is sometimes called the anomalous density, and

\begin{equation}
\psi ^{\dagger }\psi ^{2}\simeq 0
\end{equation}
Taking the expectation value of the Heisenberg equation we obtain the Gross
- Pitaievsky equation (GPE)

\begin{equation}
i\frac{\partial \Phi }{\partial t}-H\Phi +u\left| \Phi \right| ^{2}\Phi
+u\left\{ 2\Phi {\tilde{n}}+\Phi ^{\ast }{\tilde{m}}\right\} =0
\label{EOMcond}
\end{equation}
Subtracting this from the original equation Eq. (\ref{EomPsi}), and
neglecting nonlinear terms we obtain the equation for the fluctuation field 
\begin{equation}
\left[ i\frac{\partial }{\partial t}-H+2u\left| \Phi \right| ^{2}\right]
\psi \left( x\right) +u\Phi ^{2}\psi ^{\dagger }\left( x\right) =0
\label{EOMfluc}
\end{equation}
This is the basic equation we shall use here for analyzing the behavior of
the excitations (fluctuations) for a given condensate evolution. From the
evolution of the fluctuations we can then calculate the modified evolution
of the condensate. Note that only in taking the Bogoliubov approximation
(setting ${\tilde{n}}={\tilde{m}}=0$) will one get a closed equation, the
Gross-Pitaevskii equation, for the condensate. In principle these two
equations for the fluctuations (\ref{EOMfluc}) and the condensate (\ref
{EOMcond}) need be solved together in a self-consistent manner. (This is
called the `backreaction problem' first studied in external field or
cosmological particle creation problems). Here we solve only the equation
for the fluctuation field using the experimentally measured value for the
background field. As mentioned earlier this is already a (lowest order)
backreaction- modified or self-consistent solution which is an improvement
over the mean field solution obtained from the GP equation.

We next parametrize the wave functions as

\begin{equation}
\Phi =\Phi _{0}e^{-i\Theta }
\end{equation}

\begin{equation}
\psi =\psi _{0}e^{-i\Theta }
\end{equation}
where $\Phi _{0}$ and $\Theta $ are real. We seek to express the two
equations for $\Phi,\psi$ by the three equations for the three real
quantities $\Phi_0, \psi_0, \Theta$. Observe that

\begin{equation}
\nabla _{x}^{2}(Fe^{-i\Theta})=e^{-i\Theta }\left\{ \nabla
_{x}^{2}F-2i\nabla \Theta \nabla F-\left[ i\nabla ^{2}\Theta +\left( \nabla
\Theta \right) ^{2}\right] F\right\},
\end{equation}
therefore,

\begin{equation}
\frac{\partial \Theta }{\partial t}+\frac{1}{2}\left( \nabla \Theta \right)
^{2}-\frac{1}{\Phi _{0}}H\Phi _{0}+u\Phi _{0}^{2}+u\left\{ 2{\tilde{n}}+%
\mathrm{Re\;}{\tilde{m}}\right\} =0  \label{teta}
\end{equation}

\begin{equation}
\frac{\partial \Phi _{0}}{\partial t}+\nabla \Theta \nabla \Phi _{0}+\frac{1%
}{2}\left( \nabla ^{2}\Theta \right) \Phi _{0}+u\Phi _{0}\mathrm{Im\;}{%
\tilde{m}}=0  \label{phi}
\end{equation}

\begin{equation}
\left[ i\frac{\partial }{\partial t}-H+2u\Phi _{0}^{2}\right] \psi
_{0}+u\Phi _{0}^{2}\psi _{0}^{\dagger }+\psi _{0}\frac{\partial \Theta }{%
\partial t}+i\nabla \Theta \nabla \psi _{0}+\frac{1}{2}\left[ i\nabla
^{2}\Theta +\left( \nabla \Theta \right) ^{2}\right] \psi _{0}=0
\end{equation}
The third equation may be simplified by using the first

\begin{equation}
\left[ i\left( \frac{\partial }{\partial t}+\left( \nabla \Theta \right)
\nabla +\frac{1}{2}\nabla ^{2}\Theta \right) -H+\frac{1}{\Phi _{0}}\left(
H\Phi _{0}\right) +u\Phi _{0}^{2}\right] \psi _{0}+u\Phi _{0}^{2}\psi
_{0}^{\dagger }=0  \label{psi}
\end{equation}
We have also neglected terms which were nonlinear in fluctuations.

\section{Onset of collapse and scaling of $t_{collapse}$}

\subsection{The early stage}

During the first few miliseconds of evolution, we may regard the condensate
density as time independent, and the condensate phase as homogeneous (see
Appendix A for justification). We may then write the equation for the
fluctuation field Eq. (\ref{EOMfluc}) as

\begin{equation}
\left[ i\frac{\partial }{\partial t}-H+E_{0}\right] \psi _{0}+u\Phi
_{0}^{2}\left( \psi _{0}+\psi _{0}^{\dagger }\right) =0
\end{equation}

\begin{equation}
E_{0}=\frac{1}{2}\left( \omega _{z}+2\omega _{\rho }\right)
\end{equation}

To solve this equation we decompose it into a self-adjoint and an
anti-adjoint part

\begin{equation}
\psi _{0}=\xi +i\eta   \label{treintayuno}
\end{equation}
with each part satisfying an equation 
\begin{equation}
\frac{\partial \xi }{\partial t}=\left[ H-E_{0}\right] \eta 
\label{treintaydos}
\end{equation}

\begin{equation}
\frac{\partial \eta }{\partial t}+\left[ H-E_{0}-2u\Phi _{0}^{2}\right] \xi
=0.  \label{treintaytres}
\end{equation}
In the subspace orthogonal to the condensate, the operator has an inverse,
whence

\begin{equation}
\eta =\left[ H-E_{0}\right] ^{-1}\frac{\partial \xi }{\partial t},
\end{equation}
and since the trap Hamiltonian is time - independent, we have

\begin{equation}
\frac{\partial ^{2}\xi }{\partial t^{2}}+\left[ H-E_{0}\right] H_{eff}\xi =0.
\end{equation}
Here

\begin{equation}
H_{eff}=H-E_{0}-2u\Phi _{0}^{2}
\end{equation}
is the Hamiltonian for a particle moving in the potential

\begin{equation}
V\left( r\right) =\frac{1}{2}\left[ \omega _{z}^{2}z^{2}+\omega _{\rho
}^{2}\rho ^{2}-Ae^{-r^{2}}-2E_{0}\right] ,
\end{equation}
where $r^{2}=\omega _{z}z^{2}+\omega _{\rho }\rho ^{2}$ and $A=4N_{0}u/\pi
^{3/2}$. (Recall from experiment that the condensate is stable for $A\leq
16\;\left( 0.55\right) /\sqrt{\pi }=4.\,\allowbreak 96$).

Suppose $\left[ H-E_{0}\right] H_{eff}$ has eigenvectors $\psi _{K}$, with
eigenvalues $\lambda _{_{K}}$, we can expand

\begin{equation}
\xi =\sum_{\vec{n}}c_{_{K}}\left( t\right) \psi _{K}
\end{equation}
Then $c_{_{K}}\left( t\right) $ will be a superposition of two harmonic
oscillations with frequencies $\pm \sqrt{\lambda _{_{K}}}.$ To have an
unstable condensate it is necessary that at least one of the $\lambda
_{_{K}} $ is negative; the boundary of stability occurs when the lowest $%
\lambda _{_{K}},$ say $\lambda _{_{0}}$ is exactly zero. Since $\left[
H-E_{0}\right] $ has an inverse, the equation $\left[ H-E_{0}\right]
H_{eff}\psi _{0}=0$ implies $H_{eff}\psi _{0}=0$.

In conclusion, the onset of collapse corresponds to the minimum value of the
condensate density for which the effective Hamiltonian $H_{eff}$ develops a
zero mode.

\subsection{The onset of collapse}

We shall now study the spectrum of the operator $H_{eff}$. The idea is that
the low lying states, which are the most relevant ones to our discussion,
will try to keep close to where the potential is a minimum, namely, the
origin. If the width of the state is small enough, the potential will be
nearly harmonic.

One further consideration is that we are interested in the part of the
fluctuation field which remains orthogonal to the condensate, since
fluctuations along the condensate mode may be interpreted as condensate
fluctuations rather than particle loss. This is granted for all modes with
odd parity; for modes of even parity, it means that a certain combination
must be excluded. The ground state of $H_{eff}$ is certainly not orthogonal
to the condensate, since neither have nodes.

A simple but remarkably accurate way of determining whether $H_{eff}$ admits
a zero mode is to consider its expectation value $\left\langle
H_{eff}\right\rangle $ with respect to some appropriate trial state. The
first excited eigenvalue $E_{100}$ is the minimum possible expectation
value, taken with respect to any normalized state orthogonal to the ground
state. Therefore, if we may exhibit a wave function leading to a negative $%
\left\langle H_{eff}\right\rangle ,$ then necessarily $E_{100}$ is negative
as well.

Let us write

\begin{equation}
H_{eff}=H_{0}^{z}+H_{0}^{\rho }-2u\Phi _{0}^{2}-E_{0}
\end{equation}

\begin{equation}
H_{0}^{z}=\frac{p_{z}^{2}}{2}+\frac{\omega _{z}^{2}z^{2}}{2}
\end{equation}

\begin{equation}
H_{0}^{\rho }=\frac{p_{\rho }^{2}}{2}+\frac{\omega _{\rho }^{2}\rho ^{2}}{2}
\end{equation}
Recall for the condensate

\begin{equation}
\Phi _{0}^{2}=\frac{N_{0}}{\pi ^{3/2}}e^{-\left[ \omega _{z}z^{2}+\omega
_{\rho }\rho ^{2}\right] }
\end{equation}
We shall try to minimize the expectation value of $H$ with respect to a wave
function of the form

\begin{equation}
\psi _{1}^{\Omega _{z}}\left( z\right) \psi _{0}^{\Omega _{\rho }}\left(
x\right) \psi _{0}^{\Omega _{\rho }}\left( y\right)
\end{equation}
where $\psi _{0}^{\Omega }$ and $\psi _{1}^{\Omega }$ are the fundamental
and first excited states of an one-dimensional harmonic oscillator with
arbitrary frequency $\Omega $

\begin{equation}
\psi _{0}^{\Omega }=\left( \frac{\Omega }{\pi }\right) ^{1/4}e^{-\Omega
z^{2}/2}
\end{equation}

\begin{equation}
\psi _{1}^{\Omega }=\left( \frac{4\Omega }{\pi }\right) ^{1/4}\left( \Omega
^{1/2}z\right) e^{-\Omega z^{2}/2}
\end{equation}
The free part yields

\begin{equation}
\left\langle H_{0}^{z}\right\rangle =\frac{3}{4}\left( \Omega _{z}+\frac{%
\omega _{z}^{2}}{\Omega _{z}}\right)
\end{equation}

\begin{equation}
\left\langle H_{0}^{\rho }\right\rangle =\frac{1}{2}\left( \Omega _{\rho }+%
\frac{\omega _{\rho }^{2}}{\Omega _{\rho }}\right)
\end{equation}
The condensate part yields

\begin{equation}
\int d^{3}r\;\Phi _{0}^{2}\left( r\right) \left[ \psi _{1}^{\Omega
_{z}}\left( z\right) \psi _{0}^{\Omega _{\rho }}\left( x\right) \psi
_{0}^{\Omega _{\rho }}\left( y\right) \right] ^{2}=\frac{N_{0}}{\pi ^{3/2}}
\Omega _{\rho }\Omega _{z}^{3/2}\left( \omega _{z}+\Omega _{z}\right)
^{-3/2}\left( \omega _{\rho }+\Omega _{\rho }\right) ^{-1}  \label{overlap}
\end{equation}
Putting them all together

\begin{equation}
\left\langle H_{eff}\right\rangle =\frac{1}{2}\left[ \frac{3}{2}\left(
\Omega _{z}+\frac{\omega _{z}^{2}}{\Omega _{z}}\right) +\left( \Omega _{\rho
}+\frac{\omega _{\rho }^{2}}{\Omega _{\rho }}\right) -\omega _{z}-2\omega
_{\rho }-\frac{A}{\left( 1+\frac{\omega _{z}}{\Omega _{z}}\right)
^{3/2}\left( 1+\frac{\omega _{\rho }}{\Omega _{\rho }}\right) }\right]
\end{equation}
where 
\begin{equation}
A=4u\frac{N_{0}}{\pi ^{3/2}}=\frac{16}{\sqrt{\pi }}\left( N_{0}\frac{a}{%
a_{ho}}\right)
\end{equation}

If we adopt the values $\omega _{z}=1/2,$ $\omega _{\rho }=\sqrt{2},$
relevant to the JILA experiment, then we obtain $\left\langle
H_{eff}\right\rangle =0$ for the first time when $A=A_{crit}\sim 4.6,$ $%
\Omega _{z}\sim 0.78$ and $\Omega _{\rho }\sim 1.7.$ From the definition of $%
A,$ we conclude that instability will occur when

\begin{equation}
N_{0}\frac{a_{crit}}{a_{ho}}=\kappa =\frac{\sqrt{\pi }}{16}A_{crit}=0.51
\end{equation}
This result compares remarkably well with the experimental value $\kappa
=0.55,$ as well as with the theoretical estimate presented in Ref. \cite
{GTT01}. This agreement may be seen as natural, as the equations we
postulate for the fluctuations may be obtained from the linearization of the
GPE, discarding both $\mathrm{\tilde{n}}$ and $\mathrm{\tilde{m}}$. In both
calculations, the geometry of the trap plays a fundamental role.

\subsection{Scaling of $t_{collapse}$}

As we have already noted, even for condensate densities above the stability
limit, no particles is seen to be lost from the condensate during a waiting
time $t_{collapse}.$

Even in the absence of a detailed model of the condensate evolution, the
above analysis allows us to make a definite prediction of the way $%
t_{collapse}$ scales with the scattering length. The basic idea is that,
while $t_{collapse}$ depends in a complex way on several time scales, some
intrinsic to the condensate and some related to the condensate -
noncondensate interaction, up to $t_{collapse}$ the time scales intrinsic to
the condensate are very large compared to the other processes. The
nontrivial point is that even in this limit the time scales of the
noncondensate remain finite, and so they fix the scale for $t_{collapse}$
itself. Using the exponential growth of the first excited state as a
measure, we are led to the estimate $t_{collapse}\sim \varepsilon ^{-1}.$

Consider a value of $A$ close to, yet higher than, the critical one. The
fastest growing mode is the eigenvector of $\left[ H-E_{0}\right] H_{eff}$
with most negative eigenvalue, say $-\sigma .$ When $A$ takes the critical
value $A_{cr},$ $\sigma =0$. Close to the critical point, we may compute $%
\sigma $ using time independent perturbation theory, even if $\left[
H-E_{0}\right] H_{eff}$ is not Hermitian. The idea is that at the critical
value of $A$ there is a single normalized state $\left| 100\right\rangle $
orthogonal to the condensate such that $\left[ H-E_{0}\right] H_{eff}\left|
100\right\rangle =0$ and

\begin{equation}
\left\langle 100\right| \left[ H-E_{0}\right] H_{eff}\left| \psi
\right\rangle +\left\langle \psi \right| \left[ H-E_{0}\right] H_{eff}\left|
100\right\rangle =0
\end{equation}
for any state $\left| \psi \right\rangle $ orthogonal to both the condensate
and $\left| 100\right\rangle $ (otherwise, there would be states orthogonal
to the condensate with $\left\langle \left[ H-E_{0}\right]
H_{eff}\right\rangle <0,$ which is impossible). Now for $A$ slightly above
the critical value, let $\left| \sigma \right\rangle =\left|
100\right\rangle +\left| \delta \sigma \right\rangle ,$ $\left\langle
100\right| \left. \delta \sigma \right\rangle =0$ be the eigenstate
corresponding to the minimum eigenvalue $-\sigma .$ Then

\begin{equation}
-\sigma =\left\langle 100\right| \left[ H-E_{0}\right] H_{eff}\left| \sigma
\right\rangle =\left\langle 100\right| \left[ H-E_{0}\right] H_{eff}\left|
100\right\rangle +\;\mathrm{higher\;order}
\end{equation}
and so

\begin{equation}
\sigma \sim \left( \frac{A}{A_{crit}}-1\right) \left\langle \left[
H-E_{0}\right] 2u\Phi _{0}^{2}\right\rangle
\end{equation}
where the expectation value is computed at the critical point. Since we know
the form of the wave function, computing the expectation value is a simple
exercise. We find that the growing mode increases as $\exp \varepsilon t$,
where

\begin{equation}
\varepsilon =\sqrt{\sigma }\sim T\left( \frac{A}{A_{crit}}-1\right) ^{1/2}
\label{more rigorous estimate}
\end{equation}
and $T=\left\langle \left[ H-E_{0}\right] 2u\Phi _{0}^{2}\right\rangle \sim
0.6.$ Since $A/A_{crit}=a/a_{cr}$ for the given total number of atoms,

\begin{equation}
t_{collapse}=t_{crit}\left( \frac{a}{a_{cr}}-1\right) ^{-1/2}
\label{scaling}
\end{equation}

The power law Eq. (\ref{scaling}) describes with great accuracy the way $%
t_{collapse}$ scales with the scattering length; the actual prediction $%
t_{crit}\sim T^{-1}$ is correct only as an order of magnitude estimate. In
natural units, $T^{-1}$ is about $20$ms, while the best fit to the
experimental data is obtained for $t_{crit}\sim 5$ms.

We should stress that the factor of $4$ discrepancy between our prediction
and the experimental result could be easily explained away in terms of a
more complete model of condensate depletion. For example, the probability
for two particles in the condensate to collide and transfer to the growing
mode, Bose enhanced by the population in the latter, would go as $\left(
\psi ^{\dagger }\psi \right) ^{2},$ thereby yielding the factor of four in
the exponent. However, in this paper we shall keep to our goal of providing
the simplest theory which gives the best qualitative description of the Bose
Novae phenomena, hereby understood as arising from the quantum dynamics of
the fluctuations.

An analysis of collapse based on the Popov approximation equations, for a
simplified geometry, is given in \cite{Y02}. However, the time $t_{NL}\sim
\left( uN_{0}\right) ^{-1}$ \cite{TBJ00}, which is herewith identified as
characteristic of collapse, does not account for the enhancement of $%
t_{collapse}$ near the critical point.

In Fig. 3 we plot the scaling law (\ref{scaling}) (full line) derived here
and compare it with the experimental data for $N_{0}=6000$ as reported in
Refs. \cite{JILA01b,Claussen03} (small black points), the $t_{NL}\sim \left(
uN_{0}\right) ^{-1}$ prediction (suitable scaled) as given in \cite
{Y02,Claussen03} (dashed line) and the results of numerical simulations
reported in \cite{SU03} (large grey dots). While all three theoretical
predictions may be considered satisfactory, the $t_{NL}\sim \left(
uN_{0}\right) ^{-1}$ fails to describe the divergence of $t_{collapse}$ as
the critical point is approached. The results of numerical simulations
reported in \cite{SU03} based on an improved Gross-Pitaevskii equation tend
to be systematically above the experimental results. In a classical
instability, the unstable modes must grow from zero, while in a quantum
instability, they are always seeded by their own zero - point fluctuations,
which speeds up the development of the instability. Therefore, the fact that
numerical simulations based on classical instability tend to overestimate $%
t_{collapse}$ may be a further indication of the quantum origin of the
phenomenon.

\subsection{Minor role of a molecular condensate}

Having obtained some insight into the early stages of collapse, let us
discuss briefly how the present model may be modified to explicitly account
for the population of a molecular state. We refer the reader to Refs. \cite
{MSJ02b,MMH03,Mackie02} for further details.

The system of atoms and molecules at the Feshbach resonant state may be
described by a theory where the fundamental degrees of freedom are an atomic
field $\Psi \left( x\right) $ (we use the dimensionless amplitude introduced
in the previous Section) and a molecular field $A\left( x\right) .$ The
Hamiltonian takes the form

\begin{equation}
H=H_{atom}+H_{mol}+H_{be}+H_{int}
\end{equation}
where

\begin{equation}
H_{atom}=H_{trap}+H_{bkgd}.
\end{equation}
Here $H_{trap}$ given in Eq. (\ref{hachetrap}) describes the dynamics of
free atoms in the trap, $H_{bkgd}$ accounts for the background interaction
between atoms (that is, very far from resonance)

\begin{equation}
H_{bkgd}=\left( \frac{-u_{bkgd}}{2}\right) \int d^{3}x\;\Psi ^{\dagger
2}\Psi ^{2}\left( x\right)
\end{equation}
where \cite{JILA02b} (recall that our sign convention is a positive
scattering length for an attractive interaction)

\begin{equation}
u_{bkgd}=4\pi \frac{a_{bkgd}}{a_{ho}},\qquad a_{bkgd}=450\;a_{0}\Rightarrow
u_{bkgd}=0.1
\end{equation}
$H_{mol}$ describes the motion of molecules in the trap. We treat a molecule
as a particle of mass $2M$ (where $M$ is the mass of the atom) subject to a
potential $2V$ (where $V$ is the trap potential seen by the atoms) \cite
{WFHRH00}, so the frequency of oscillations in the trap is the same for
atoms and molecules. $H_{be}$ accounts for the energy difference between a
molecule and two atoms 
\begin{equation}
H_{be}=-\varepsilon \left( B\right) \int d^{3}x\;A^{\dagger }A\left( x\right)
\end{equation}
Plots of the binding energy $\varepsilon \left( B\right) $ as a function of
the applied field $B$ are given in \cite{KH02,JILA03}. Finally, $H_{int}$
accounts for the formation and dissociation of molecules

\begin{equation}
H_{int}=\Gamma \int d^{3}x\;\left( A^{\dagger }\Psi ^{2}+\Psi ^{\dagger
2}A\right) \left( x\right)
\end{equation}

The binding energy $\varepsilon \left( B\right) $ is the crucial input
parameter in the model. $\varepsilon \left( B\right) =0$ at the resonance
field $B=B_{peak}=155$G \cite{JILA03} and increases with larger $B$'s. At
values of $B\sim 156$G typical of the Ramsey fringes experiment \cite
{JILA02b} we already have binding energies of the order of $10$KHz; this is
much larger than the $100$Hz typical of motion in the trap, and therefore $%
H_{mol}$ is negligible (in the condensate collapse experiment \cite{JILA01b}
typical fields where $B\sim 167$G, and $\varepsilon \left( B\right) $ was
much larger). Under this approximation, the $full$ Heisenberg equation of
motion for the molecular field

\begin{equation}
i\frac{\partial }{\partial t}A=-\varepsilon \left( B\right) A+\Gamma \Psi
^{2}
\end{equation}
may be solved analytically

\begin{equation}
A\left( r,t\right) =A\left( r,0\right) e^{i\varepsilon t}-i\Gamma
\int_{0}^{t}dt\prime \;e^{i\varepsilon \left( t-t^{\prime}\right) }\Psi
^{2}\left( r,t^{\prime}\right)
\end{equation}
$If$ the atomic field is slowly varying in the scale of $\varepsilon ^{-1},$
then the integral is dominated by the upper limit, and simplifies to

\begin{equation}
A\left( r,t\right) =A_{free}\left( r,t\right) +\frac{\Gamma }{\varepsilon }%
\Psi ^{2}\left( r,t\right)
\end{equation}
The free part decouples from the atomic condensate, and the effect of the
driven part is to introduce an effective interaction among atoms, so that $%
u_{bkgd}$ in $H_{bkgd}$ is replaced by

\begin{equation}
u_{eff}=u_{bkgd}-\frac{2\Gamma ^{2}}{\varepsilon \left( B\right) }
\end{equation}
which is the familiar pattern which makes it possible to tune the scattering
length. The effective scattering length vanishes at $B=B_{zero}=165.75$G 
\cite{JILA03} (see Fig. 1).

We conclude that under this approximation, the number of molecules -
disregarding those already present independently of the atomic condensate -
is

\begin{equation}
N_{mol}\sim \left( \frac{\Gamma }{\varepsilon }\right) ^{2}N_{atom}^{2}=%
\frac{u_{bkgd}}{2\varepsilon \left( B\right) }\left( 1-\frac{u_{eff}}{%
u_{bkgd}}\right) N_{atom}^{2}
\end{equation}
If $N_{atom}\sim 16000$ and $u_{eff}\sim -u_{bkgd}$ \cite{JILA02b}, then the
molecular condensate becomes important when $\varepsilon \left( B\right)
\sim u_{bkgd}N_{atom}\omega _{ho}= 128$kHz. This regime was achieved in \cite
{JILA02b}. Under these conditions we cannot rely on the above approximation,
but rather we must solve the model. This yields atom - molecule oscillations
with a frequency set by $\varepsilon \left( B\right) .$

In contrast, in the condensate collapse experiment \cite{JILA01b} the
magnetic field was driven $above$ $B_{zero}$, with typical values $B=167$G.
For these fields, the molecular binding energy is close to $12$MHz (see \cite
{JILA03,Kok03}), or $15\;10^{4}\;\omega _{0}$ Therefore $N_{mol}\sim
0.005N_{atom}\sim \allowbreak 80$. This is less than the number of atoms
even in the earliest measured jets (see \cite{JILA01b} and below),
suggesting that no major role is played by molecular recombination. This
conclusion is consistent with the fact that no oscillations in the number of
particles in the condensate are seen (see \cite{JILA01b} and below). It is
interesting to observe that the numerical simulations presented in Ref. \cite
{MMH03} assume a set of parameters chosen to display the effect most
clearly, not attempting to be realistic.

This leaves open the possibility that the approximation may break down
because the atomic field is not slowly varying with respect to $\varepsilon
^{-1}.$ For example, the time it takes to change the magnetic field from $%
B_{zero}$ to $B_{evolve}$ is probably of the order of $10\mu $s$%
=8\;10^{-4}\omega _{ho}^{-1}$ \cite{JILA02b}. This is a short time compared
to $\varepsilon ^{-1}$ under the conditions of the Ramsey fringes experiment 
\cite{JILA02b}, but a rather easy stroll in the condensate collapse
experiment \cite{JILA01b}. The next scale in which we expect the condensate
to react (which is also of the order of the inverse atomic chemical
potential) is $t_{NL}$ \cite{TBJ00,Y02}, but this is of the order of
milliseconds and therefore way too long.

The conclusion is that formation or dissociation of molecules, under the
conditions of the condensate collapse experiment \cite{JILA01b}, is unlikely
to have played a central role. Moreover, in models where the molecular state
is explicitly included, it is still necessary to account for the quantum
dynamics of the atomic field. This is done, for example, in Ref. \cite{MMH03}
by tracking the fluctuation two-point functions along with the condensates,
or in \cite{Mackie02} by providing independent equations for the anomalous
atom pair amplitude. Resonance of molecular field with the condensate alone,
though remarkably successful in the later experiments, cannot account for
the phenomena of this experiment. For the Bose Novae experiment, the
dominant qualitative features are determined by the dynamical vacuum
squeezing mechanism proposed here; including a molecular field can only
provide a minor quantitative improvement.

\section{Evolution of fluctuations: Bursts and Jets}

In the previous Section, we were mostly concerned with identifying the
factors which can render the condensate unstable. In this Section, we shall
consider the quantum evolution of fluctuations, as a test field riding on
the collapsing condensate extracted from experiment. The basic dynamical
equations are, from before, Eqs. (\ref{treintayuno}), (\ref{treintaydos})
and (\ref{treintaytres}), where now $\Phi _{0}^{2}$ is assumed to have both
space and time dependence. The initial state is defined by the condition
that $u=0$ for $t<0;$ we shall take it to be the particle vacuum $\left|
0\right\rangle ,$ which is defined by $\psi \left( x,0\right) \left|
0\right\rangle =0$ everywhere. Rather than seeking a rigorous solution, we
wish to understand the structure of the problem in order to display the
extent to which the phenomenology of condensate collapse is determined by
fundamental (yet simple) quantum dynamical effects such as parametric
amplification and particle creation. With this goal in mind, we shall make
some physically motivated simplifications.

We have already seen that it is possible to introduce a mode decomposition
of the $\xi $ operator based on the eigenfunctions of $\left[ H-E_{0}\right]
H_{eff}.$ This shows that the preconceived notion that fluctuations will
react only to the local state of the condensate is flawed; the relevant
modes are nonlocal and can sample conditions over a large portion of the
trap.

For short wavelengths $\lambda $, we expect these eigenfunctions will
approach trap eigenmodes, since $H\sim \lambda ^{-2}>>2u\Phi _{0}^{2}$. Also
the fact that particles in bursts are seen to oscillate with the trap
frequencies \cite{JILA01b} suggests that their dynamics is determined by the
trap Hamiltonian (see below). Of course, the trap eigenmodes would also be
eigenfunctions of $H_{eff}$ if this operator and $H$ commuted. Now, since

\begin{equation}
\left[ H,H_{eff}\right] =u\left\{ \nabla ^{2}\Phi _{0}^{2}+2\nabla \Phi
_{0}^{2}\nabla \right\}
\end{equation}
we see that we may disregard the conmutator when the condensate is slowly
varying with respect to the relevant modes. Early enough in the collapse,
the typical scale for the condensate is $a_{ho},$ and this condition will
hold for almost every mode. Thus we shall make the approximation of assuming
a homogeneous condensate. The amplitude of the homogeneous condensate will
be chosen to enforce the onset of instability. Since we expect the lowest
trap mode to become unstable when $N=N_{cr}=\kappa /a$ (in units where $%
a_{ho}=1$), we approximate

\begin{equation}
2u\Phi _{0}^{2}\equiv \left( \frac{a}{\kappa }\right) \omega _{z}N_{0}\left(
t\right)  \label{homocond}
\end{equation}
where $N_{0}\left( t\right) $ is the instantaneous total number of particles
in the condensate. For constant $N_{0},$ this yields $\varepsilon \sim
\omega _{z}\left( a/a_{cr}-1\right) ^{1/2}$ for the time constant of the
growing mode, which is close to the more rigorous estimate Eq. (\ref{more
rigorous estimate}) above.

In practice, $\kappa ^{-1}$ in Eq.(\ref{homocond}) is a measure of the
overlap between the condensate and the excitation modes, as in Eq.(\ref
{overlap}) above. Therefore, the approximation may be improved by adjusting $%
\kappa $ according to the range of modes where it will be applied. We shall
adopt this practice in what follows.

\subsection{Evolution of fluctuations up to $t_{collapse}$}

We now proceed with the quantitative analysis of Bose Novae. We assume a
given evolution extracted from experiment for the number of particles in the
condensate and analyze the evolution of fluctuations treated as a test field
on this dynamic background.

Concretely, we shall consider evolutions where the number of particles in
the condensate remains constant from the time $t=0$ when the scattering
length is switched to a negative value up to some time $t_{collapse}$, and
decays exponentially from then on (see Fig. 4). In the actual experiments
the condensate does not evaporate completely, but the number of particles in
the remnant is much smaller than the initial number of atoms in the
condensate, and is therefore negligible at the early times when the
approximation of a homogeneous condensate is valid.

Let us begin by considering the evolution up to the time of collapse $%
t_{collapse}$, or the waiting period. Let $\bar{N}_{0}$ be the initial
number of particles in the condensate, and $a_{cr}=\kappa /\bar{N}_{0}$ the
corresponding critical scattering length. Trap eigenfunctions are labelled
by a string of quantum numbers $\vec{n}=\left( n_{z},n_{x},n_{y}\right) .$
The eigenvalues of the trap Hamiltonian are (with the zero energy already
subtracted) $E_{\vec{n}}=\vec{n}\vec{\omega}$ where $\vec{\omega}=\left(
\omega _{z},\omega _{\rho },\omega _{\rho }\right) $ and $\vec{n}\vec{\omega}%
=\omega _{z}n_{z}+\omega _{\rho }\left( n_{x}+n_{y}\right) $. We choose the
eigenfunctions $\psi _{\vec{n}}$ to be real.

Let us expand

\begin{equation}
\psi =\sum_{\vec{n}}a_{\vec{n}}\left( t\right) \psi_{\vec{n}}\left( r\right)
\end{equation}

\begin{equation}
\xi =\sum_{\vec{n}}c_{\vec{n}}\left( t\right) \psi_{\vec{n}}
\end{equation}

\begin{equation}
\eta =\sum_{\vec{n}}b_{\vec{n}}\left( t\right) \psi_{\vec{n}}
\end{equation}
Then

\begin{equation}
c_{\vec{n}}=\frac{1}{2}\left( a_{\vec{n}}+a_{\vec{n}}^{\dagger }\right)
\end{equation}

\begin{equation}
b_{\vec{n}}=\frac{1}{2i}\left( a_{\vec{n}}-a_{\vec{n}}^{\dagger }\right)
\end{equation}

\begin{equation}
\frac{d}{dt}b_{\vec{n}}=-\left[ E_{\vec{n}}-\left( \frac{a}{a_{cr}}\right)
\omega _{z}\right] c_{\vec{n}}
\end{equation}

\begin{equation}
\frac{d}{dt}c_{\vec{n}}=E_{\vec{n}}b_{\vec{n}}
\end{equation}
We see that there are two kinds of modes, stable (oscillatory, or thawed)
modes if $E_{\vec{n}}>\left( \frac{a}{a_{cr}}\right) \omega _{z},$ and
unstable (growing, or frozen) modes if not. We shall consider each kind in
turn.

\subsection{Stable modes and bursts}

First consider the case when $E_{\vec{n}}>\left( \frac{a}{a_{cr}}\right)
\omega _{z}.$ We get

\begin{equation}
c_{\vec{n}}\left( t\right) =\frac{1}{2}\left( a_{\vec{n}}+a_{\vec{n}%
}^{\dagger }\right) \left( 0\right) \cos \omega _{\vec{n}}t+\frac{1}{2i\chi_{%
\vec{n}}}\left( a_{\vec{n}}-a_{\vec{n}}^{\dagger }\right) \left( 0\right)
\sin \omega _{\vec{n}}t
\end{equation}

\begin{equation}
b_{\vec{n}}\left( t\right) =\frac{1}{2i}\left( a_{\vec{n}}-a_{\vec{n}%
}^{\dagger }\right) \left( 0\right) \cos \omega _{\vec{n}}t-\frac{1}{2}\chi
_{\vec{n}}\left( a_{\vec{n}}+a_{\vec{n}}^{\dagger }\right) \left( 0\right)
\sin \omega _{\vec{n}}t
\end{equation}
where

\begin{equation}
\omega _{\vec{n}}=\sqrt{E_{\vec{n}}\left[ E_{\vec{n}}-\left( \frac{a}{a_{cr}}%
\right) \omega _{z}\right] }=\chi _{\vec{n}}E_{\vec{n}}
\end{equation}

\begin{equation}
\chi _{\vec{n}}=\sqrt{1-\left( \frac{a}{a_{cr}}\right) \frac{\omega _{z}}{E_{%
\vec{n}}}}.
\end{equation}
So

\begin{equation}
a_{\vec{n}}\left( t\right) =f_{\vec{n}}\left( t\right) a_{\vec{n}}\left(
0\right) +g_{\vec{n}}\left( t\right) a_{\vec{n}}^{\dagger }\left( 0\right),
\end{equation}
where

\begin{equation}
f_{\vec{n}}\left( t\right) =\cos \omega _{\vec{n}}t+\left[ 2-\left( \frac{a}{%
a_{cr}}\right) \frac{\omega _{z}}{E_{\vec{n}}}\right] \frac{\sin \omega _{%
\vec{n}}t}{2i\chi _{\vec{n}}},
\end{equation}

\begin{equation}
g_{\vec{n}}\left( t\right) =-\left( \frac{a}{a_{cr}}\right) \frac{\omega _{z}%
}{E_{\vec{n}}}\frac{\sin \omega _{\vec{n}}t}{2i\chi _{\vec{n}}}.
\end{equation}
Observe that

\begin{equation}
\left| f_{\vec{n}}\left( t\right) \right| ^{2}-\left| g_{\vec{n}}\left(
t\right) \right| ^{2}=\cos ^{2}\omega _{\vec{n}}t+\left\{ \left[ 2-\left( 
\frac{a}{a_{cr}}\right) \frac{\omega _{z}}{E_{\vec{n}}}\right] ^{2}-\left( 
\frac{a}{a_{cr}}\frac{\omega _{z}}{E_{\vec{n}}}\right) ^{2}\right\} \frac{%
\sin ^{2}\omega _{\vec{n}}t}{4\chi _{\vec{n}}^{2}}=1
\end{equation}
as required by the commutation relations. Although we assume vacuum initial
conditions, these modes do not remain empty. The density

\begin{eqnarray}
\mathrm{\tilde{n}}\left( r,t\right) &=&\sum_{\vec{n}}\psi _{_{\vec{n}%
}}^{2}\left( r\right) \left\langle a_{\vec{n}}^{\dagger }\left( t\right) a_{%
\vec{n}}\left( t\right) \right\rangle  \nonumber \\
&=&\sum_{\vec{n}}\psi _{\vec{n}}^{2}\left( r\right) \left| g_{\vec{n}}\left(
t\right) \right| ^{2}=\frac{1}{8}\left( \frac{a}{a_{cr}}\right) ^{2}\omega
_{z}^{2}\sum_{\vec{n}}\psi _{\vec{n}}^{2}\left( r\right) \frac{\sin
^{2}\omega _{\vec{n}}t}{\omega _{\vec{n}}^{2}}
\end{eqnarray}
We see that the density has a constant term and an oscillatory term. In our
view this oscillatory term is responsible for the appearance of ``bursts''
of particles oscillating within the trap. This point is worthy of some
elaboration, as it will make clearer the contrast with the unstable modes to
be discussed below.

Let us approximate the amplitudes $\psi _{\vec{n}}\left( r\right) $ by their
WKB forms

\begin{equation}
\psi _{\vec{n}}\left( r\right) =\prod_{i=1-3}\psi _{_{i}}\left( r_{i}\right)
\end{equation}

\begin{equation}
\psi _{_{i}}\left( r_{i}\right) =\frac{1}{\sqrt{p_{i}}}\cos \left[
S_{i}-n_{i}\frac{\pi }{2}\right]
\end{equation}
where

\begin{equation}
S_{i}=\int_{0}^{r_{i}}dx\;p_{i}\left( x\right)
\end{equation}

\begin{equation}
p_{i}=\sqrt{2n_{i}\omega _{i}-\omega _{i}^{2}r_{i}^{2}}
\end{equation}
After expanding all trigonometric functions, we see that the oscillatory
part is a linear combination of terms of the form

\begin{equation}
\exp \left\{ 2i\left[ \sum S_{i}-\omega _{\vec{n}}t\right] \right\}
\end{equation}
in all sign combinations. If we look at the density at a given point and
time $\left( r,t\right) ,$ we see that the modes which contribute
effectively are those for which (in the limit $\chi _{\vec{n}}\rightarrow 1,$
$E_{\vec{n}}\gg \omega _{z}$)

\begin{equation}
\omega _{i}\left[ \int_{0}^{r_{i}}\frac{dx}{p_{i}\left( x\right) }-t\right]
=0
\end{equation}
This equation describes a particle which moves along a classical trajectory
in the trap potential, starting from the origin at $t=0$ with momentum $%
p_{i}\left( 0\right) $. If we include the phase shift in computing the
saddle point, there is a time delay, but it is the same for all particles.

We conclude that the oscillatory part of the density describes a swarm of
particles moving along classical trajectories in the trap potential. These
trajectories return to the origin at multiples of the inverse trap
frequencies, thus producing a surge in the local density, which becomes
large enough to be seen by destructive absorption imaging. These particles
constitute the so-called \textbf{bursts} observed in the Bose - Nova
experiment.

\subsection{Unstable modes and jets}

Let us now consider the opposite case $E_{\vec{n}} \le \left( \frac{a}{a_{cr}%
}\right) \omega _{z}.$ Formally we may obtain the relevant formulae from the
last section with the replacement $\chi _{\vec{n}}=i\vartheta _{\vec{n}},$ $%
\omega _{\vec{n}}=i\sigma _{\vec{n}}$, thus transforming $\cos \omega _{\vec{%
n}}t\rightarrow \cosh \sigma _{\vec{n}}t$ and $\sin \omega _{\vec{n}%
}t\rightarrow i\sinh \sigma _{\vec{n}}t.$ We get

\begin{equation}
c_{\vec{n}}\left( t\right) =\frac{1}{2}\left( a_{\vec{n}}+a_{\vec{n}%
}^{\dagger }\right) \left( 0\right) \cosh \sigma _{\vec{n}}t+\frac{1} {%
2i\vartheta _{\vec{n}}}\left( a_{\vec{n}}-a_{\vec{n}}^{\dagger }\right)
\left( 0\right) \sinh \sigma _{\vec{n}}t
\end{equation}

\begin{equation}
b_{\vec{n}}\left( t\right) =\frac{1}{2i}\left( a_{\vec{n}}-a_{\vec{n}%
}^{\dagger }\right) \left( 0\right) \cosh \sigma _{\vec{n}}t+\frac{1}{2}%
\vartheta _{\vec{n}}\left( a_{\vec{n}}+a_{\vec{n}}^{\dagger }\right) \left(
0\right) \sinh \sigma _{\vec{n}}t
\end{equation}
So again

\begin{equation}
a_{\vec{n}}\left( t\right) =f_{\vec{n}}\left( t\right) a_{\vec{n}}\left(
0\right) +g_{\vec{n}}\left( t\right) a_{\vec{n}}^{\dagger }\left( 0\right)
\end{equation}
But now

\begin{equation}
f_{\vec{n}}\left( t\right) =\cosh \sigma _{\vec{n}}t+\frac{1}{2i\vartheta_{%
\vec{n}}}\left( 1-\vartheta _{\vec{n}}^{2}\right) \sinh \sigma _{\vec{n}}t
\end{equation}

\begin{equation}
g_{\vec{n}}\left( t\right) =\frac{-1}{2i\vartheta _{\vec{n}}}\left(
1+\vartheta _{_{\vec{n}}}^{2}\right) \sinh \sigma _{\vec{n}}t=\frac{i}{%
2\vartheta _{\vec{n}}}\left( \frac{a}{a_{cr}}\right) \frac{\omega _{z}}{E_{%
\vec{n}}} \sinh \sigma _{\vec{n}}t
\end{equation}

Physically the difference is huge. In the first place, the density is
growing exponentially. But unlike the previous case, there is no oscillatory
component. While the actual number of particles is increasing, there are no
surges in the density caused by the sudden constructive interference of many
modes. In particular these particles do not contribute to the central peak
in the density distribution. In this sense, they can not be seen by
destructive absorption imaging. Because these particles do not oscillate in
the trap, in the sense above, we say these modes are frozen in the same
sense used in theories of cosmological structure formation, i.e., that
fluctuations in an evolving universe are said to freeze upon leaving the
horizon \cite{STA86}.

However, these modes come alive at $\tau _{evolve}$, when the scattering
length is set to zero. Now they become ordinary trap modes, and oscillate in
the trap in the same way as the burst described above. To the observer, they
appear as a new injection of particles from the core of the condensate,
which makes up the so-called `\textbf{jets}'. The sudden activation of a
frozen mode by turning off the particle - particle interaction may be
described as a ``thaw''.

Observe that in this picture several conspicuous features of jets become
obvious. Jets may only appear the turn - off time $\tau _{evolve}$ is
earlier than the formation of the remnant. Once the condensate is stable
again, there are no more frozen modes to thaw. On the other hand, jets will
appear (as observed) for $\tau _{evolve}<t_{collapse}$, when the condensate
has not yet shed any particles. Also jets must be less energetic than
bursts, since they are composed of lower modes. Because of their relative
weakness, treating them as test particles riding on the dynamic condensate
as is the approximation adopted here (`test field') works better for jets
than for bursts. A more accurate depiction of the dynamics of bursts may
require consideration of the backreaction of the fluctuations on the
condensate.

\subsection{Beyond $t_{collapse}:$ thawing}

The main physics we presented in the last Section is that the amplitude for
a normal mode in the excitation (fluctuations) of the condensate evolves as
a harmonic oscillator with a complex frequency dictated by the dynamics of
the condensate. Parametric amplification (or, in a quantum optics,
squeezing) of the fluctuations populates the modes above the condensate,
even from a vacuum state originally, manifesting as bursts or jets. Thus in
our view the salient features of Bose - Novae, the specific phenomena
arising from the collapse of a BEC formed from a dilute Bose gas arises from
the squeezing of (vacuum) fluctuations by the condensate dynamics.

Let us now discuss the behavior of fluctuations beyond $t_{collapse}$, when
the number of particles in the condensate, and therefore the instantaneous
frequency of the excited modes, become time dependent. As in the previous
Section, we shall assume nevertheless that the condensate remains
homogeneous, thereby confining ourselves to the early stages of collapse.

For concreteness, let us assume that the number of particles in the
condensate remains constant $N_{0}=\bar{N}_{0}=16,000$ up to $%
t=t_{collapse}=3$ms, and then decays exponentially $N_{0}\left( t\right) =%
\bar{N}_{0}\mathrm{exp}\left( -\left( t-t_{collapse}\right) /\tau \right) $,
with $\tau =6$ms (see Fig. 4). The mean frequency of the trap $\omega =80$Hz
corresponds to a temperature $\hbar \omega /k_{B}=52.5$nK. The actual trap
frequencies reported in Ref. \cite{JILA01b} $\omega _{radial}=110$ Hz and $%
\omega _{axial}=42.7$ Hz correspond to temperatures $\allowbreak
28.\,\allowbreak 012$ $\ $(axial) and $\allowbreak 72.\,\allowbreak 162$
(radial) nK. These are relatively high with respect to the sample
temperature of $3$ nK. For this reason, the initial number of particles
above the condensate is negligible, and we may assume that we are dealing
with particle creation from effectively the vacuum. This is the rationale
behind calling this process `squeezing of the vacuum'. We shall assume the
scattering length is brought to $a=36a_{0}$. We shall approximate the effect
of the condensate on the fluctuations by a constant level shift as in Eq.(%
\ref{homocond}). The best fit to the experimental data is obtained for $%
\kappa =0.46,$ which is satisfactorily close to the experimental value of $%
0.55.$

Shifting the origin of time to $t_{collapse}$ for simplicity, we write $%
N_{0}\left( t\right) =\bar{N}_{0}\mathrm{exp}\left( -t/\tau \right) .$ After
expanding in trap eigenmodes as in the previous Section, we obtain the
equations

\begin{equation}
\frac{d}{dt}b_{\vec{n}}=-\left[ E_{\vec{n}}-\left( \frac{a\omega _{z}}{\bar{a%
}}\right) \mathrm{exp}\left( -t/\tau \right) \right] c_{\vec{n}}
\end{equation}

\begin{equation}
\frac{d}{dt}c_{\vec{n}}=E_{\vec{n}}b_{\vec{n}}
\end{equation}
Therefore

\begin{equation}
\frac{d^{2}}{dt^{2}}c_{\vec{n}}+E_{\vec{n}}\left[ E_{_{\vec{n}}}-\left( 
\frac{a\omega _{z}}{\bar{a}}\right) \mathrm{exp}\left( -t/\tau \right)
\right] c_{\vec{n}}=0
\end{equation}
This equation clearly displays the two kinds of behavior described above (an
exact solution is provided in Appendix B). If $E_{\vec{n}}>\left( \frac{%
a\omega _{z}}{\bar{a}}\right) ,$ the mode is always oscillatory. If $E_{_{%
\vec{n}}}<\left( \frac{a\omega _{z}}{\bar{a}}\right) ,$ the mode is frozen
at $t_{collapse},$ but thaws when $\mathrm{exp}\left( -t/\tau \right) \sim
E_{\vec{n}}\bar{a}/a\omega _{z}$ (see Fig. 5)$.$ During the frozen period,
the modes are amplified, but they only contribute to bursts after thaw. If
the evolution is interrupted while still frozen, they appear as a jet.

We therefore conclude that the number of particles $N_{jet}$ in a jet at
time $\tau _{evolve}$ is essentially the total number of particles in all
frozen modes at that time. If we write as before 
\begin{equation}
a_{\vec{n}}\left( t\right) =f_{\vec{n}}\left( t\right) a_{\vec{n}}\left(
0\right) +g_{\vec{n}}\left( t\right) a_{\vec{n}}^{\dagger }\left( 0\right)
\end{equation}
then

\begin{equation}
N_{jet}\left( t\right) =\sum_{E_{\vec{n}}\leq \left( \frac{a\omega _{z}}{%
\bar{a}}\right) \mathrm{exp}\left( -t/\tau \right) }\left| g_{\vec{n}}\left(
t\right) ^{2}\right|
\end{equation}
This is plotted in Fig 6, from the exact solution in Appendix B, with the
parameters given above, and compared to the corresponding results as
reported in \cite{JILA01b}. We see that the agreement is excellent at early
times (up to about $6$ms). For latter times, the model overestimates the jet
population. This is due to the fact that, by not considering the shrinking
of the condensate, we are overestimating the overlap between the condensate
and the fluctuations, thus delaying the thaw. It nevertheless reproduces the
overall slope of particle number with $\tau _{evolve},$ which is quite
remarkable considering the simplicity of the model.

The see-saw pattern follows from the discreteness of the modes. Modes thaw
at discrete times $t_{k}^{\ast }.$ Between one and the next, occupation
numbers of the unstable modes are increasing, and so the number of particles
in the jet is a growing function of $\tau _{evolve}$. When the next
stabilization occurs, the particles in the now stable mode no longer
contribute to later jets, and the particle number in the jet decreases by a
finite amount. This pattern accounts for the fact that a jet from a later $%
\tau _{evolve}$ may be stronger than earlier ones, and also for the large
variation in the number of particles in jets with similar values of $\tau
_{evolve}$. It should also be remembered that we are computing the expected
number of particles, but that, in the highly squeezed state which results
from the frozen period, the fluctuations in particle number are comparable
to the mean number itself.

\section{Conclusion}

In this paper, we have applied insights from the quantum field theory of
particle creation and structure formation in cosmological spacetimes and the
theory of second order phase transitions to a specific scenario of
controlled collapse of a Bose-Einstein condensate, the so-called Bose Novae
phenomena. We have described these phenomena as resulting from particle
creation from the vacuum, induced by the time dependent condensate. This
time dependence squeezes and amplifies the field operator describing
excitations above the condensate. A key concept in our analysis borrowed
from theories of cosmological structure formation is the drastic difference
in the physical effects of frozen versus oscillatory modes: those whose
physical frequencies are higher than the collapse rate oscillate and are
rather impervious to the condensate, while those below (frozen modes) grow
in time and get amplified, in a way similar to the growth of fluctuations
during spinodal decomposition. As the condensate stabilizes and the collapse
rate decreases the frozen modes begin to thaw. The appearance of oscillatory
modes (in second quantized language) is described as particle creation
appearing in jets and bursts, as described in detail above.

In order to focus on the key ideas we have adopted a number of simplifying
assumptions. We take the condensate evolution as a given input from the
experiments, rather than deriving it from fully self-consistent equations.
We have treated excitations within the Popov approximation, which improves
on the Hartree approach but is known to break down as the number of
particles above the condensate increases. We have neglected the coupling
between different excitation modes, considering only the coupling of each to
the condensate.

These simplifications render certain aspects of the problem more amenable to
others because they are rather insensitive to the assumptions. The scaling
of $t_{collapse}$ is shown to depend on the behavior of a few modes setting
the characteristic time scale of the problem - therefore the prediction is
not affected by the underestimation of the coupling to other modes.

Even within these simplications, we have obtained good quantitative
predictions for the onset of instability, the scaling of the waiting time $%
t_{collapse}$ (when the condensate implosion really begins after the
inversion of the scattering length) with the scattering length, and also for
the number of particles in a jet as a function of $\tau _{evolve}$, when the
interaction between atoms is switched off.

Another success of the model is to provide a simple explanation for the
widely different appearance of bursts and jets. As remarked earlier, jets
may only appear if the turn - off time $\tau _{evolve}$ is earlier than the
formation of the remnant, because once the condensate is stable again, there
are no more frozen modes to thaw, but, on the other hand, jets will appear
for $\tau _{evolve}<t_{collapse}$, when the condensate has not yet shed any
particles. Also jets must be less energetic than bursts, since they are
composed of lower modes.

Considering the success these simple ideas and light calculations brought
about we believe our approach might have captured the essence of the physics
behind these phenomena. The physical paradigms we used in bringing forth
these ideas also suggest that understanding the basic mechanism of important
processes in cosmology, critical dynamics, and Bose-Einstein condensation
may share more than a superficial ground.\newline

\textbf{Acknowledgement} We first learned about this neat experiment from
Bill Phillips who also introduced us to Keith Burnett whose `industrial
strength' colloquium at Maryland and discussions afterwards provided us with
the background information about Bose Novae. Our colleagues Ted Jacobson and
Stefano Liberati have been looking for Hawking radiation and analog models
of gravity in BEC, thus providing our group with a supercool ambience. We
thank them all for enticing us to join in the BEC fun. The audiences in EC's
seminar at UMD and BLH's seminar at NIST reporting on these results are to
be thanked for their questions. We are obliged to E. Donley and S.
Kokkelmans for their prompt and informative responses to our questions on
BECs and for communicating key unpublished data. EC also acknowledges
discussions with Eric Bolda. This research and EC's visits to UMD are
supported in part by a NSF grant PHY98-00967, a NIST grant and by CONICET,
UBA, Fundacion Antorchas and ANPCyT under grant PICT99 03-05229.

\section{Appendix A: the classical GPE}

In this appendix we shall consider the purely classical GPE, meaning that we
shall disregard any back reaction from fluctuations. The conclusion of this
analysis is that for short times of the order of $1$ms, it is possible to
consider the modulus of the condensate as time independent, and its phase as
space independent. This observation will lead to a substantial
simplification of the equations for the fluctuations in this regime.

Let us begin from the equations

\begin{equation}
\frac{\partial \Theta }{\partial t}+\frac{1}{2}\left( \nabla \Theta \right)
^{2}-\frac{1}{\Phi _{0}}H\Phi _{0}+u\Phi _{0}^{2}=0
\end{equation}

\begin{equation}
\frac{\partial \Phi _{0}}{\partial t}+\nabla \Theta \nabla \Phi _{0}+\frac{1%
}{2}\left( \nabla ^{2}\Theta \right) \Phi _{0}=0
\end{equation}
It is suggestive to rewrite these equations in the following way. The
operator

\begin{equation}
D_{t}=\frac{\partial }{\partial t}+\left( \nabla \Theta \right) \nabla
\end{equation}
looks a lot as a material derivative with respect to a fluid flowing at each
point with velocity $v=\nabla \Theta $. It seems natural to change from the
eulerian coordinates $x$ to Lagrangian coordinates $q.$ That is, for given $%
q $ we define the function $x=x\left( q,t\right) $ as the solution to the
system

\begin{equation}
\left. \frac{\partial x\left( q,t\right) }{\partial t}\right| _{q}=\nabla
_{x}\Theta \left( x\left( q,t\right) ,t\right) ,\qquad x\left( q,0\right) =q
\end{equation}
Observe that

\begin{equation}
D_{t}=\left. \frac{\partial }{\partial t}\right| _{q}
\end{equation}
It is also convenient to define the fields

\begin{equation}
J_{\alpha }^{i}=\left. \frac{\partial x^{i}}{\partial q^{\alpha }}\right|
\end{equation}
and their inverses $J_{i}^{\alpha }.$ At any given time, the $q^{\alpha }$ \
are a set of curvilinear coordinates, with metric

\begin{equation}
ds^{2}=g_{\alpha \beta }dq^{\alpha }dq^{\beta }
\end{equation}

\begin{equation}
g_{\alpha \beta }=\delta _{ij}J_{\alpha }^{i}J_{\beta }^{j}
\end{equation}
The wave function transforms as a scalar under this coordinate change. We
also define

\begin{equation}
J=\det \;J_{\alpha }^{i}
\end{equation}
Observe that

\begin{equation}
\left. \frac{\partial J}{\partial t}\right| _{q}=J\;J_{i}^{\alpha }\frac{%
\partial ^{2}\Theta }{\partial q^{\alpha }\partial x^{i}}=J\;\nabla
^{2}\Theta.
\end{equation}
The volume element is $\sqrt{g}d^{3}q,$ $g=\det g_{\alpha \beta }=J^{2},$
and the Laplacian

\begin{equation}
\frac{1}{\sqrt{g}}\partial _{\alpha }\sqrt{g}g^{\alpha \beta }\partial
_{\beta }
\end{equation}
This suggests writing

\begin{equation}
\Phi _{0}=\frac{\mathbf{\Phi }_{0}}{\sqrt{J}}
\end{equation}
To get the equation

\begin{equation}
\frac{\partial \mathbf{\Phi }_{0}}{\partial t}=0
\end{equation}
Writing the dynamics this way, the role of the phases is hidden in the time
dependent curvilinear coordinate system, since we get

\begin{equation}
\mathbf{\Phi }_{0}=\mathbf{\Phi }_{0}\left( q\right) =\frac{\sqrt{N_{0}}}{%
\pi ^{3/4}}e^{-q^{2}/2}
\end{equation}
where we are using the fact that initially the condensate corresponds to a
noninteracting gas.

To be definite, consider a situation with $N_{0}=16000$ and $a=30\;a_{0}.$
We obtain

\begin{equation}
N_{0}u=4\pi \frac{a}{a_{ho}}N_{0}=104
\end{equation}
Since we expect that, at least initially, $H\mathbf{\Phi }_{0}\sim 3\mathbf{%
\Phi }_{0}/2$, the dynamics of the phase near the origin is dominated by the 
$u\mathbf{\Phi }_{0}^{2}$ term. Concretely, observe that

\begin{equation}
\left. \frac{\partial \Theta }{\partial t}\right| _{x}=\left. \frac{\partial
\Theta }{\partial t}\right| _{q}-\left( \nabla \Theta \right) ^{2}
\end{equation}
So the equation for the phase reads

\begin{equation}
\frac{\partial \Theta }{\partial t}-\frac{1}{2}\left( \nabla \Theta \right)
^{2}-\frac{\sqrt{J}}{\mathbf{\Phi }_{0}}H\frac{\mathbf{\Phi }_{0}}{\sqrt{J}}+%
\frac{u}{\sqrt{J}}\mathbf{\Phi }_{0}^{2}=0
\end{equation}
Approximate

\begin{equation}
\Theta \left( q,t\right) =\mu _{0}-t\left[ u\mathbf{\Phi }_{0}^{2}\left(
q\right) -\frac{3}{2}\right]
\end{equation}
Observe that

\begin{equation}
\nabla \Theta =2tuq\mathbf{\Phi }_{0}^{2}\left( q\right)
\end{equation}
So

\begin{equation}
\frac{1}{2}\left( \nabla \Theta \right) ^{2}=\left( 2t^{2}uq^{2}\mathbf{\Phi 
}_{0}^{2}\left( q\right) \right) u\mathbf{\Phi }_{0}^{2}\left( q\right)
\end{equation}
This term is therefore negligible against $u\mathbf{\Phi }_{0}^{2}$ if $q\gg
1$ or $q\leq \left( t\sqrt{N_{0}u}\right) ^{-1};$ it is small everywhere if $%
t\leq \left( \sqrt{N_{0}u}\right) ^{-1}.$ This is enough to justify the
formal procedures in the main body of this paper.

The equation which defines the coordinate change from $x$ to $q$ becomes

\begin{equation}
\left. \frac{\partial x\left( q,t\right) }{\partial t}\right| _{q}=2tuq%
\mathbf{\Phi }_{0}^{2}\left( q\right) ,\qquad x\left( q,0\right) =q,
\end{equation}
with solution

\begin{equation}
x=q\left[ 1+t^{2}u\mathbf{\Phi }_{0}^{2}\left( q\right) \right]
\end{equation}
So again, we see that during the first millisecond, we may approximate $x=q$
\ everywhere. This establishes the conclusions anticipated at the beginning
of this section.

\section{Appendix B: exact solutions of the mode equation in the dynamical
case}

In this Appendix we shall derive closed form solutions for the evolution
equations for quantum fluctuations after $t_{collapse}$

\begin{equation}
\frac{d}{dt}b_{\vec{n}}=-\left[ E_{\vec{n}}-\left( \frac{a\omega _{z}}{\bar{a%
}}\right) \mathrm{exp}\left( -t/\tau \right) \right] c_{\vec{n}}
\end{equation}

\begin{equation}
\frac{d}{dt}c_{\vec{n}}=E_{\vec{n}}b_{\vec{n}}
\end{equation}
Therefore

\begin{equation}
\frac{d^{2}}{dt^{2}}c_{\vec{n}}+E_{\vec{n}}\left[ E_{_{\vec{n}}}-\left( 
\frac{a\omega _{z}}{\bar{a}}\right) \mathrm{exp}\left( -t/\tau \right)
\right] c_{\vec{n}}=0
\end{equation}
Call

\begin{equation}
\zeta =\zeta _{0}e^{\left( -t/2\tau \right) }
\end{equation}
Therefore

\begin{equation}
\frac{d}{dt}=\frac{-\zeta }{2\tau }\frac{d}{d\zeta }
\end{equation}
and

\begin{equation}
\frac{1}{4\tau ^{2}}\zeta \frac{d}{d\zeta }\zeta \frac{d}{d\zeta }c_{_{\vec{n%
}}}+E_{\vec{n}}\left[ E_{\vec{n}}-\left( \frac{a\omega _{z}}{\bar{a}}\right)
\left( \frac{\zeta }{\zeta _{0}}\right) ^{2}\right] c_{\vec{n}}=0
\end{equation}
so

\begin{equation}
\frac{d^{2}}{d\zeta ^{2}}c_{\vec{n}}+\frac{1}{\zeta }\frac{d}{d\zeta }c_{%
\vec{n}}-\left( \frac{4\tau ^{2}a\omega _{z}E_{\vec{n}}}{\bar{a}\zeta
_{0}^{2}}-\frac{4\tau ^{2}E_{\vec{n}}^{2}}{\zeta ^{2}}\right) c_{_{\vec{n}%
}}=0
\end{equation}

Choose

\begin{equation}
\zeta _{0}^{2}=\frac{4\tau ^{2}a\omega _{z}E_{\vec{n}}}{\bar{a}}
\end{equation}
to find (recall that the $c_{\vec{n}}$ are self-adjoint)

\begin{equation}
c_{\vec{n}}=\alpha _{\vec{n}}I_{2i\tau E_{\vec{n}}}\left( \zeta \right)
+\alpha _{\vec{n}}^{\dagger }I_{-2i\tau E_{\vec{n}}}\left( \zeta \right)
\end{equation}
We see the two basic behaviors discussed in the previous section. For $%
t<<\tau ,$ $\zeta >>1,$ the Bessel functions behave as real exponentials.
When $t\rightarrow \infty ,$ $\zeta \rightarrow 0$ the modes oscillate with
frequency $\pm E_{\vec{n}}.$ Modes with $E_{\vec{n}}>a\omega _{z}/\bar{a}$
are never frozen. Lower modes start frozen at $t=0,$ but thaw when $\zeta
\sim 2\tau E_{\vec{n}},$ or

\begin{equation}
e^{\left( -t/\tau \right) }\sim \frac{\bar{a}E_{\vec{n}}}{a\omega _{z}}
\end{equation}

The $b_{\vec{n}}$ coefficients are given by

\begin{equation}
b_{\vec{n}}=\frac{1}{E_{\vec{n}}}\frac{dc_{\vec{n}}}{dt}=\frac{-\zeta }{2E_{%
\vec{n}}\tau }\frac{dc_{\vec{n}}}{d\zeta }
\end{equation}
The $c_{\vec{n}}$ and $b_{\vec{n}}$ coefficients must be continuous, so

\begin{equation}
\alpha _{\vec{n}}I_{2i\tau E_{\vec{n}}}\left( \zeta _{0}\right) +\alpha _{%
\vec{n}}^{\dagger }I_{-2i\tau E_{\vec{n}}}\left( \zeta _{0}\right) =\bar{c}_{%
\vec{n}}
\end{equation}

\begin{equation}
\frac{-\zeta _{0}}{2E_{\vec{n}}\tau }\left[ \alpha _{\vec{n}}\frac{d}{d\zeta 
}I_{2i\tau E_{\vec{n}}}\left( \zeta _{0}\right) +\alpha _{\vec{n}}^{\dagger }%
\frac{d}{d\zeta }I_{-2i\tau E_{\vec{n}}}\left( \zeta _{0}\right) \right] =%
\bar{b}_{\vec{n}}
\end{equation}
where $\bar{c}_{\vec{n}}$ and $\bar{b}_{\vec{n}}$ are the result of the
evolution up to $t_{collapse},$ namely

\begin{equation}
\bar{c}_{\vec{n}}=\frac{1}{2}\left( a_{\vec{n}}+a_{\vec{n}}^{\dagger
}\right) \left( 0\right) \cosh \sigma _{\vec{n}}t_{collapse}+\frac{1}{2i}%
\frac{\left( a_{\vec{n}}-a_{\vec{n}}^{\dagger }\right) \left( 0\right) }{%
\vartheta _{\vec{n}}}\sinh \sigma _{\vec{n}}t_{collapse}
\end{equation}

\begin{equation}
\bar{b}_{\vec{n}}=\frac{1}{2i}\left( a_{\vec{n}}-a_{\vec{n}}^{\dagger
}\right) \left( 0\right) \cosh \sigma _{\vec{n}}t_{collapse}+\frac{1}{2}%
\vartheta _{\vec{n}}\left( a_{\vec{n}}+a_{\vec{n}}^{\dagger }\right) \left(
0\right) \sinh \sigma _{\vec{n}}t_{collapse}
\end{equation}

Using the Wronskian

\begin{equation}
I_{2i\tau E_{\vec{n}}}\left( \zeta _{0}\right) \frac{d}{d\zeta }I_{-2i\tau
E_{\vec{n}}}\left( \zeta _{0}\right) -I_{-2i\tau E_{_{\vec{n}}}}\left( \zeta
_{0}\right) \frac{d}{d\zeta }I_{2i\tau E_{\vec{n}}}\left( \zeta _{0}\right)
=-2i\frac{\sinh 2\pi \tau E_{\vec{n}}}{\pi \zeta _{0}}
\end{equation}
we get

\begin{equation}
\alpha _{\vec{n}}=\frac{i\pi \zeta _{0}}{2\sinh 2\pi \tau E_{\vec{n}}}\left[ 
\frac{d}{d\zeta }I_{-2i\tau E_{\vec{n}}}\left( \zeta _{0}\right) \bar{c}_{%
\vec{n}}+I_{-2i\tau E_{\vec{n}}}\left( \zeta _{0}\right) \frac{2E_{\vec{n}%
}\tau }{\zeta _{0}}\bar{b}_{\vec{n}}\right]
\end{equation}

Call

\begin{equation}
W\left( \zeta ,\zeta _{0}\right) =\frac{i\pi \zeta _{0}}{2\sinh 2\pi \tau E_{%
\vec{n}}}\left[ I_{2i\tau E_{\vec{n}}}\left( \zeta \right) I_{-2i\tau E_{%
\vec{n}}}\left( \zeta _{0}\right) -I_{-2i\tau E_{_{\vec{n}}}}\left( \zeta
\right) I_{2i\tau E_{\vec{n}}}\left( \zeta _{0}\right) \right]
\end{equation}

\begin{equation}
W_{1}\left( \zeta ,\zeta _{0}\right) =\frac{i\pi \zeta _{0}}{2\sinh 2\pi
\tau E_{\vec{n}}}\left[ I_{2i\tau E_{\vec{n}}}\left( \zeta \right) \frac{d}{%
d\zeta }I_{-2i\tau E_{\vec{n}}}\left( \zeta _{0}\right) -I_{-2i\tau E_{\vec{n%
}}}\left( \zeta \right) \frac{d}{d\zeta }I_{2i\tau E_{\vec{n}}}\left( \zeta
_{0}\right) \right]
\end{equation}

\begin{equation}
W_{2}\left( \zeta ,\zeta _{0}\right) =\frac{d}{d\zeta }W\left( \zeta ,\zeta
_{0}\right)
\end{equation}

\begin{equation}
W_{3}\left( \zeta ,\zeta _{0}\right) =\frac{d}{d\zeta }W_{1}\left( \zeta
,\zeta _{0}\right)
\end{equation}
Then

\begin{equation}
c_{\vec{n}}=W_{1}\left( \zeta ,\zeta _{0}\right) \bar{c}_{\vec{n}}+\frac{2E_{%
\vec{n}}\tau }{\zeta _{0}}W\left( \zeta ,\zeta _{0}\right) \bar{b}_{\vec{n}}
\end{equation}

\begin{equation}
b_{\vec{n}}=\frac{-\zeta }{2E_{\vec{n}}\tau }\left[ W_{3}\left( \zeta ,\zeta
_{0}\right) \bar{c}_{\vec{n}}+\frac{2E_{\vec{n}}\tau }{\zeta _{0}}%
W_{2}\left( \zeta ,\zeta _{0}\right) \bar{b}_{\vec{n}}\right]
\end{equation}

\begin{equation}
\bar{c}_{\vec{n}}=\frac{1}{2}\left( \bar{a}_{\vec{n}}+\bar{a}_{\vec{n}%
}^{\dagger }\right) ;\qquad \bar{b}_{\vec{n}}=\frac{1}{2i}\left( \bar{a}_{%
\vec{n}}-\bar{a}_{\vec{n}}^{\dagger }\right)
\end{equation}

\begin{eqnarray}
a_{\vec{n}}\left( t\right) &=&W_{1}\left( \zeta ,\zeta _{0}\right) \bar{c}_{%
\vec{n}}+\frac{2E_{\vec{n}}\tau }{\zeta _{0}}W\left( \zeta ,\zeta
_{0}\right) \bar{b}_{\vec{n}} \\
&&-\frac{i\zeta }{2E_{\vec{n}}\tau }\left[ W_{3}\left( \zeta ,\zeta
_{0}\right) \bar{c}_{\vec{n}}+\frac{2E_{\vec{n}}\tau }{\zeta _{0}}%
W_{2}\left( \zeta ,\zeta _{0}\right) \bar{b}_{\vec{n}}\right]
\end{eqnarray}
So, writing

\begin{equation}
\bar{a}_{\vec{n}}=\bar{f}_{\vec{n}}a_{\vec{n}}\left( 0\right) +\bar{g}_{\vec{%
n}}a_{\vec{n}}^{\dagger }\left( 0\right)
\end{equation}
Then

\begin{equation}
a_{\vec{n}}\left( t\right) =f_{\vec{n}}\left( t\right) a_{\vec{n}}\left(
0\right) +g_{\vec{n}}\left( t\right) a_{\vec{n}}^{\dagger }\left( 0\right)
\end{equation}

\begin{eqnarray}
f_{\vec{n}}\left( t\right) &=&\frac{1}{2}W_{1}\left( \zeta ,\zeta
_{0}\right) \left( \bar{f}_{\vec{n}}+\bar{g}_{\vec{n}}^{\ast }\right) 
\nonumber \\
&&-\frac{iE_{\vec{n}}\tau }{\zeta _{0}}W\left( \zeta ,\zeta _{0}\right)
\left( \bar{f}_{\vec{n}}-\bar{g}_{\vec{n}}^{\ast }\right)  \nonumber \\
&&-\frac{i\zeta }{4E_{\vec{n}}\tau }W_{3}\left( \zeta ,\zeta _{0}\right)
\left( \bar{f}_{\vec{n}}+\bar{g}_{\vec{n}}^{\ast }\right)  \nonumber \\
&&-\frac{\zeta }{2\zeta _{0}}W_{2}\left( \zeta ,\zeta _{0}\right) \left( 
\bar{f}_{\vec{n}}-\bar{g}_{\vec{n}}^{\ast }\right)
\end{eqnarray}

\begin{eqnarray}
g_{\vec{n}}\left( t\right) &=&\frac{1}{2}W_{1}\left( \zeta ,\zeta
_{0}\right) \left( \bar{f}_{\vec{n}}^{\ast }+\bar{g}_{\vec{n}}\right) 
\nonumber \\
&&+\frac{iE_{\vec{n}}\tau }{\zeta _{0}}W\left( \zeta ,\zeta _{0}\right)
\left( \bar{f}_{\vec{n}}^{\ast }-\bar{g}_{\vec{n}}\right)  \nonumber \\
&&-\frac{i\zeta }{4E_{\vec{n}}\tau }W_{3}\left( \zeta ,\zeta _{0}\right)
\left( \bar{f}_{\vec{n}}^{\ast }+\bar{g}_{\vec{n}}\right)  \nonumber \\
&&+\frac{\zeta }{2\zeta _{0}}W_{2}\left( \zeta ,\zeta _{0}\right) \left( 
\bar{f}_{\vec{n}}^{\ast }-\bar{g}_{\vec{n}}\right)
\end{eqnarray}
This result is used to build the plot in Fig. 6.

\newpage

\section{Figure Captions}

\subsubsection{Fig. 1}

Effective scattering length as a function of applied magnetic field (unlike
in the body of the paper, we follow here the usual sign convention). The
scattering lenght is measured in multiples of Bohr radius $a_{0}=0.529$\ $%
10^{-10}$m. The magnetic field is measured in Gauss. In the Bose Novae
experiment \cite{JILA01b,Claussen03}, the condensate is prepared at $B_{zero}
$ and then evolved to a $larger$ field. In later experiments \cite
{JILA02a,JILA02b} the field was turned to a $lower$ value, close to $B_{peak}
$ for a short time, and somewhere between $B_{peak}$ and $B_{zero}$ for the
duration of the experiment.

\subsubsection{Fig. 2}

Qualitative evolution of the effective scattering length as a function of
time in the Bose Novae experiment \cite{JILA01b,Claussen03} (unlike in the
body of the paper, we follow here the usual sign convention). The scattering
lenght in measured in multiples of Bohr radius $a_{0}=0.529$\ $10^{-10}$m,
time is measured in ms. The condensate is prepared at $a=0,$ and then the
effective interaction is made attractive. After a time $\tau _{evolve},$ the
interaction is made repulsive, which allows the condensate to expand and
aids visualization.

\subsubsection{Fig. 3}

Plot of $t_{collapse}$ (in ms) against $a_{collapse}$ (in multiples of Bohr
radius $a_{0}=0.529$\ $10^{-10}$m). We plot the scaling law Eq. (\ref
{scaling}) (full line) and compare it against the experimental data for $%
N_{0}=6000$ as reported in Refs. \cite{JILA01b,Claussen03} (small black
dots), the $t_{NL}\sim \left( uN_{0}\right) ^{-1}$ prediction (suitably
scaled) as given in \cite{Y02,Claussen03} (dashed line) and the results of
numerical simulations reported in \cite{SU03} (large grey dots). While all
three theoretical predictions may be considered satisfactory, the $%
t_{NL}\sim \left( uN_{0}\right) ^{-1}$ fails to describe the divergence of $%
t_{collapse}$ as the critical point is approached. The results of numerical
simulations based on an improved Gross-Pitaievskii equation tend to be
systematically above the experimental results$.$ In a classical instability,
the unstable modes must grow from zero, while in a quantum instability, they
are always seeded by their own zero - point fluctuations, which speeds up
the development of the instability. Therefore, the fact that numerical
simulations tend to overestimate $t_{collapse}$ may be a further indication
of the quantum origin of the phenomenon.

\subsubsection{Fig. 4}

The evolution for the number of particles in the condensate $N\left(
t\right) $ as a function of time (measured in ms) assumed for the
calculation of the number of particles in a jet. For comparison, we have
superimposed the data from Fig. 1b of \cite{JILA01b}. The agreement is
satisfactory for our purposes, as we shall not consider the evolution beyond 
$\tau _{evolve}=12$ms nor the formation of a remnant.

\subsubsection{Fig 5}

A schematic depiction of the evolution of the proper frequencies of
different modes in time, both in arbitrary units. Short wavelength modes
remain stable throughout, but the sudden change in proper frequency when the
interaction is switched on induces particle creation in them. These
particles are perceived as bursts. Long wavelength modes actually become
unstable (or frozen) until the condensate has lost enough atoms. During this
period they are amplified. When they become stable again, they are seen as a
secondary emission, or jet.

\subsubsection{Fig. 6}

The evolution of the number of particles in a jet as a function of the time $%
t_{evolve}$ (measured in ms), for $\bar{N}_{0}=16,000,$ $\omega _{radial}=110
$ Hz, $\omega _{axial}=42.7$ Hz, $a=36a_{0},$ and $\kappa =0.46$ (see Eq. (%
\ref{homocond})). The evolution of $N\left( t\right) $ in time is depicted
in Fig. 4. To provide a visual reference we have superimposed the data in
fig. 6 of \cite{JILA01b}. The discrepancy between theory and experiment
beyond $t_{evolve}=6$ms may be attributed to an overestimation of the
condensate-noncondensate coupling in neglecting the change in the shape of
the condensate.

\newpage

\begin{figure}[tbp]
\includegraphics[height=8cm]{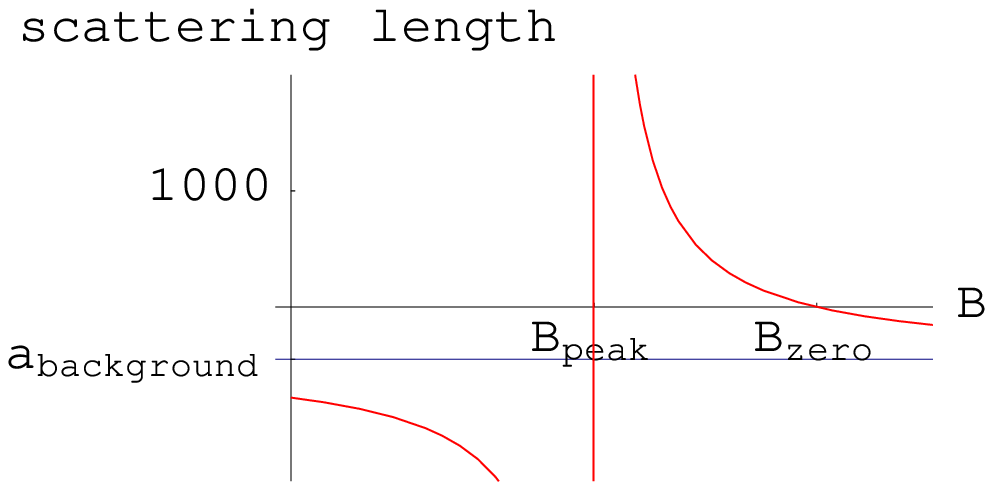}
\caption{{}}
\end{figure}

\begin{figure}[tbp]
\includegraphics[height=8cm]{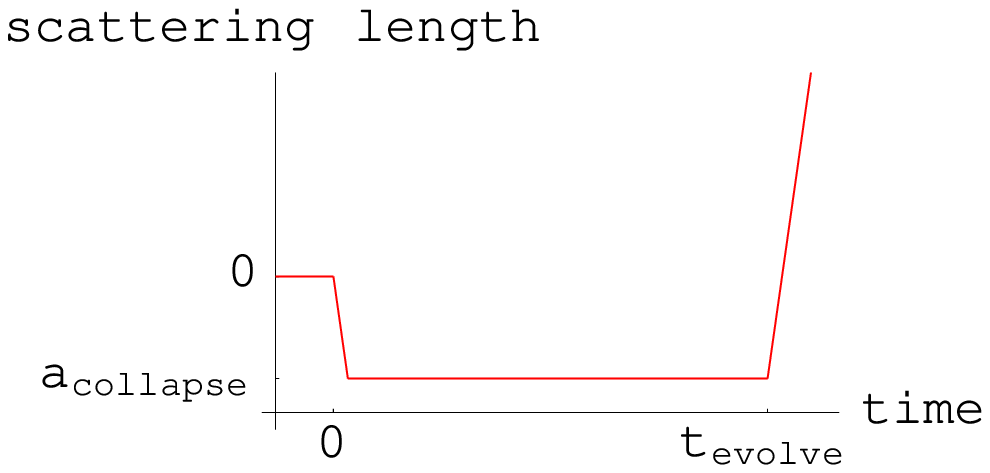}
\caption{}
\end{figure}

\begin{figure}[tbp]
\includegraphics[height=8cm]{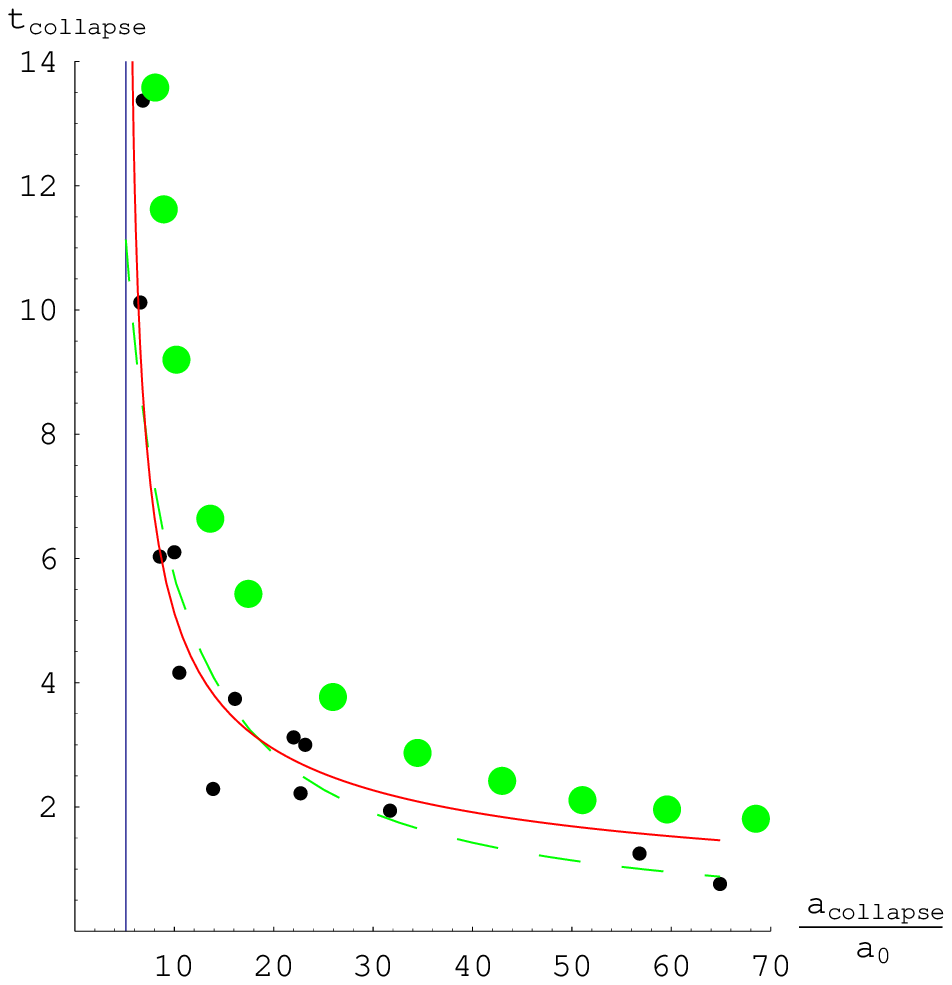}
\caption{}
\end{figure}

\begin{figure}[tbp]
\includegraphics[height=8cm]{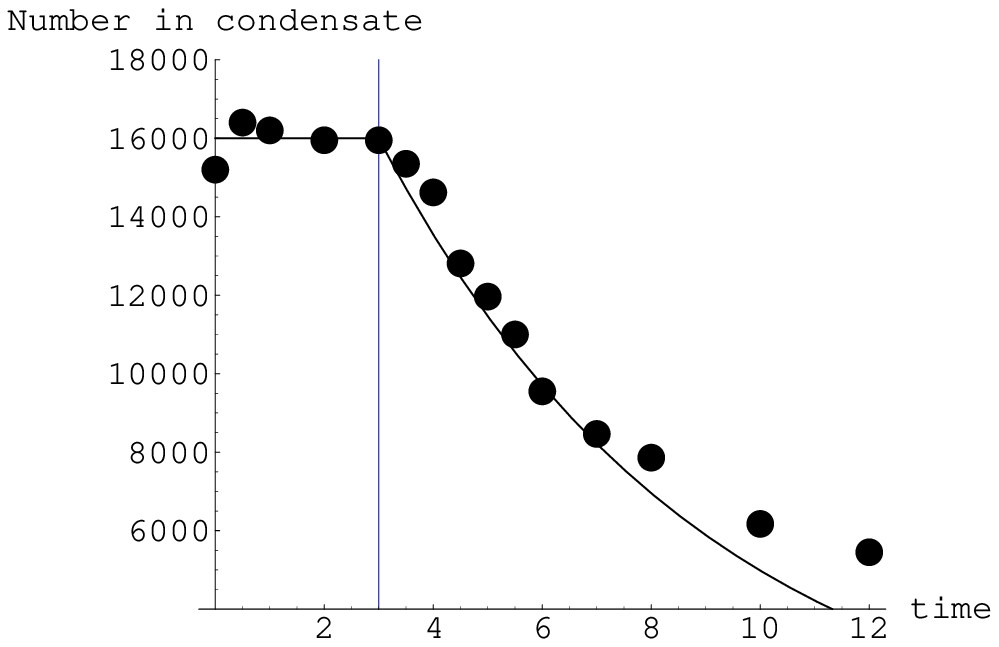}
\caption{{}}
\end{figure}

\begin{figure}[tbp]
\includegraphics[height=8cm]{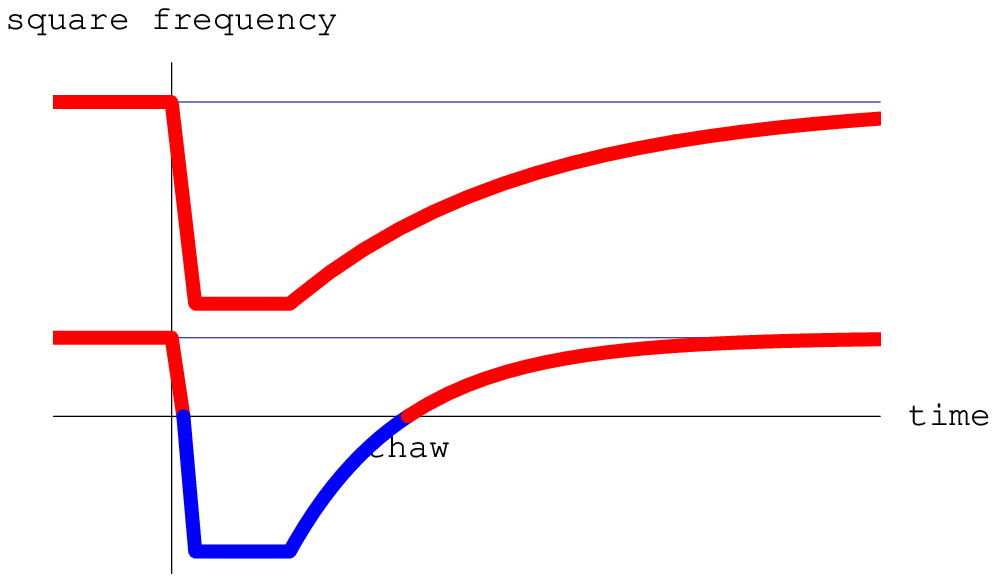}
\caption{}
\end{figure}

\begin{figure}[tbp]
\includegraphics[height=8cm]{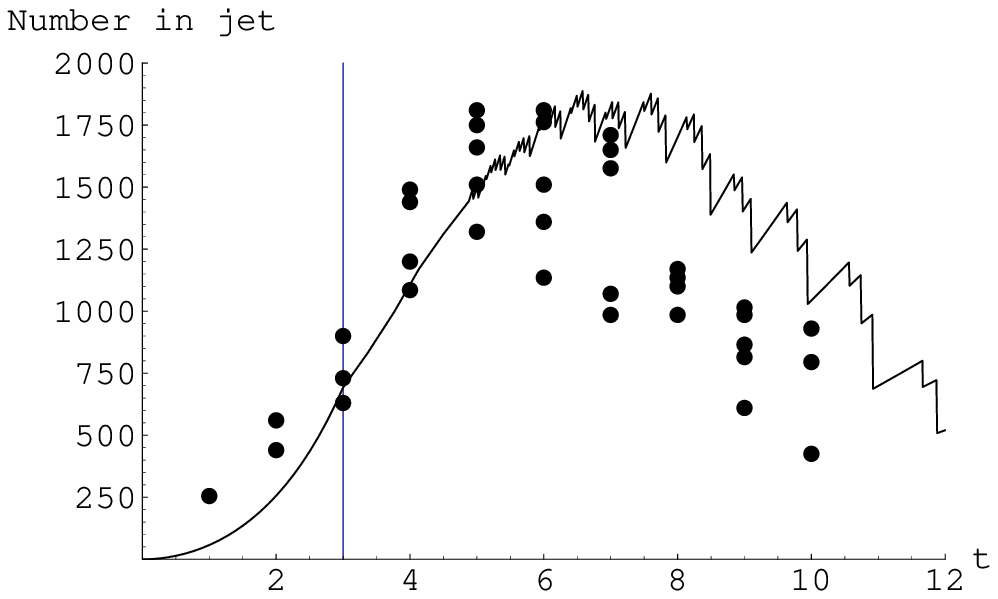}
\caption{}
\end{figure}

\end{document}